\documentclass[twocolumn,showpacs,preprintnumbers,a4,prb]{revtex4}
\usepackage{graphicx}
\usepackage{dcolumn}
\usepackage{amsmath}
\usepackage{amssymb}
\usepackage{bm}

\begin{document}

\title{Synergy of exchange bias with superconductivity in ferromagnetic-superconducting
layered hybrids: the influence of in-plane and out-of-plane
magnetic order on superconductivity}

\author{D. Stamopoulos,\cite{cor} E. Manios, and M. Pissas}

\affiliation{Institute of Materials Science, NCSR "Demokritos",
153-10, Aghia Paraskevi, Athens, Greece.}
\date{\today}

\begin{abstract}
It is generally believed that superconductivity and magnetism are
two antagonistic long-range phenomena. However, as it was
preliminarily highlighted in D. Stamopoulos {\it et al.} [Phys.
Rev. B {\bf 75}, 014501 (2007)] and extensively studied in this
work under specific circumstances these phenomena instead of being
detrimental to each other may even become cooperative so that
their synergy may promote the superconducting properties of a
hybrid structure.

Here, we have studied systematically the magnetic and transport
behavior of such {\it exchange biased} hybrids that are comprised
of ferromagnetic (FM) Ni$_{80}$Fe$_{20}$ and low-T$_c$
superconducting (SC) Nb for the case where the magnetic field is
applied {\it parallel} to the specimens. Two structures have been
studied: FM-SC-FM trilayers (TLs) and FM-SC bilayers (BLs).
Detailed magnetization data on the {\it longitudinal} and {\it
transverse} magnetic components are presented for both the {\it
normal} and {\it superconducting} states. These data are compared
to systematic transport measurements including I-V
characteristics. The comparison of the exchange biased BLs and TLs
that are studied here with the plain ones studied in D.
Stamopoulos {\it et al.} [Phys. Rev. B {\bf 75}, 184504 (2007)]
enable us to reveal an underlying parameter that may falsify the
interpretation of the transport properties of relevant FM-SC-FM
TLs and FM-SC BLs investigated in the recent literature: the
underlying mechanism motivating the extreme magnetoresistance
peaks in the TLs relates to the suppression of superconductivity
mainly due to the magnetic coupling of the two FM layers as the
{\it out-of-plane} rotation of their magnetizations takes place
across the coercive field where stray fields emerge in their whole
surface owing to the multidomain magnetic state that they acquire.
The relative {\it in-plane} magnetization configuration of the
outer FM layers exerts a secondary contribution on the SC
interlayer. Since the exchange bias directly controls the in-plane
magnetic order it also controls the out-of-plane rotation of the
FMs' magnetizations so that the magnetoresistance peaks may be
tuned at will.

All the contradictory experimental data reported in the recent
literature are fairly discussed under the light of our results;
based on a specific prerequisite we propose a phenomenological
stray-fields mechanism that explains efficiently the evolution of
magnetoresistance effect in TLs. Our experiments not only point
out the need for a new theoretical treatment of FM-SC hybrids but
also direct us toward the design of efficient supercurrent-switch
elemental devices.
\end{abstract}

\pacs{74.45.+c, 74.78.Fk, 74.78.Db}

\maketitle

\pagebreak

\section{Introduction}

In recent years a new class of devices efficient for the control
of current flow has emerged. Since unlikely to conventional
electronics in this new class it is the electron's spin (than its
charge degree of freedom) that is manipulated these devices are
called spin valves. Various types of spin valves have been
proposed. In all cases the basic core consists of a trilayer (TL)
which hosts two ferromagnetic (FM) outer electrodes with an
interlayer which can be either
metallic,\cite{Monsma95,Dieny94,Baibich88} insulating
\cite{Gider98,Mathon01,Yuasa04,Djayaprawira05} or superconducting
(SC).\cite{Gu02,Potenza05,Moraru06,Moraru06B,Pena05,Rusanov06,Visani07,Singh07,Buzdin99,Tagirov99,EschrigTL,Jin07}

The so-called superconducting spin valve was introduced
theoretically in Refs.\onlinecite{Buzdin99,Tagirov99}. It is based
on a FM-SC-FM TL where the nucleation of superconductivity can be
manipulated by the relative in-plane magnetization orientation of
the outer FM layers. Gu et al. \cite{Gu02} were the first to
report on the experimental realization of a
[Ni$_{82}$Fe$_{18}$-Cu$_{0.47}$Ni$_{0.53}$]/Nb/[Cu$_{0.47}$Ni$_{0.53}$-Ni$_{82}$Fe$_{18}$]
spin valve. Soon after, Potenza et al., \cite{Potenza05} and
Moraru et al. \cite{Moraru06,Moraru06B} confirmed the results that
were reported in Ref.\onlinecite{Gu02}. In these works
\cite{Gu02,Potenza05,Moraru06,Moraru06B} the exchange bias was
employed in order to "pin" the magnetization of the one FM layer
so that the relative magnetic configuration of the outer FM layers
to be controlled. In agreement to the theoretical propositions
\cite{Buzdin99,Tagirov99} it was observed
\cite{Gu02,Potenza05,Moraru06,Moraru06B} that when the relative
in-plane magnetization configuration of the two FM layers was
parallel [antiparallel] the resistive transition of the SC was
placed at lower [higher] temperatures. Pe\~{n}a et al.,
\cite{Pena05} and Rusanov et al. \cite{Rusanov06} have studied
La$_{0.7}$Ca$_{0.3}$MnO$_3$-YBa$_2$Cu$_3$O$_7$-La$_{0.7}$Ca$_{0.3}$MnO$_3$
and Ni$_{80}$Fe$_{20}$-Nb-Ni$_{80}$Fe$_{20}$ TLs, respectively.
Both works \cite{Pena05,Rusanov06} reported that the antiparallel
in-plane magnetization configuration of the FM layers suppresses
superconductivity when compared to the parallel one. Very
recently, Visani et al. \cite{Visani07} studied
La$_{0.7}$Ca$_{0.3}$MnO$_3$-YBa$_2$Cu$_3$O$_7$-La$_{0.7}$Ca$_{0.3}$MnO$_3$
TLs and confirmed the results that were originally reported in
Ref. \onlinecite{Pena05}. While the works mentioned above
\cite{Gu02,Potenza05,Moraru06,Moraru06B,Pena05,Rusanov06,Visani07}
referred to the case where the magnetic field was applied parallel
to the TLs (parallel-field configuration), Singh et
al.\cite{Singh07} reported on the case where it was applied normal
to the specimens (normal-field configuration) for Co-Pt/Nb/Co-Pt
TLs having perpendicular magnetic anisotropy and confirmed
Refs.\onlinecite{Pena05,Rusanov06,Visani07}. It should be stressed
that in Refs.\onlinecite{Pena05,Rusanov06,Visani07,Singh07} the
exchange bias was {\it not} used for the control of the relative
in-plane magnetic orientation of the two FM layers. Given that the
results of Refs.\onlinecite{Gu02,Potenza05,Moraru06,Moraru06B},
where the exchange bias was employed, are the opposite to the ones
of Refs.\onlinecite{Pena05,Rusanov06,Visani07,Singh07}, where this
mechanism was {\it not} employed, it is natural to suppose that,
except for the relative in-plane magnetization configuration of
the outer FM layers, it is the presence of exchange bias that
probably has influenced the properties of the SC interlayer
altering drastically the experimental results obtained in
Refs.\onlinecite{Gu02,Potenza05,Moraru06,Moraru06B}.

Indeed, detailed magnetization data that were presented in
Refs.\onlinecite{StamopoulosPRB05,StamopoulosSST06,StamopoulosPRB06}
for the parallel-field configuration in hybrids constructed of
multilayer (ML)
[La$_{0.33}$Ca$_{0.67}$MnO$_{3}$/La$_{0.60}$Ca$_{0.40}$MnO$_{3}]_{15}$
and SC Nb revealed that the exchange bias mechanism clearly
influences the SC's magnetic behavior. More specifically, the
magnetization results presented in
Refs.\onlinecite{StamopoulosPRB05,StamopoulosSST06} revealed an
effective ferromagnetic coupling between the ML and the adjacent
SC that in microscopic level was compatible with recent
theoretical proposals on the formation of spin-triplet
superconductivity that were presented in
Refs.\onlinecite{EschrigTL,Volkov03,Bergeret01,Bergeret04B,EschrigTP1,EschrigTP2}
(see also Refs.\onlinecite{VolkovLanl1,VolkovLanl2}). Also, in
Refs.\onlinecite{StamopoulosSST06,StamopoulosPRB06} it was shown
that superconductivity was enhanced significantly when the ML was
in a state of almost zero bulk magnetization where a multidomain
configuration was acquired. The magnetization results of
Refs.\onlinecite{StamopoulosSST06,StamopoulosPRB06} supported the
recent theoretical proposals on the so-called "domain-wall
superconductivity" of
Refs.\onlinecite{Buzdin84,Buzdin03R,Buzdin03,Eschrig05,Eschrig05L,Maleki06,BuzdinReview,Buzdinnew}
suggesting that when the superconducting pairs have the
opportunity to sample different directions of the exchange field
the imposed pair-breaking effect is only minimum. On the other
hand, when these results are discussed in a macroscopic level
apart from the exchange interaction also the electromagnetic
mechanism \cite{BuzdinReview,Buzdinnew} related to the magnetic
stray fields that enter the SC alongside the interface with the ML
could motivate equally well the experimental effects observed in
Refs.\onlinecite{StamopoulosPRB05,StamopoulosSST06,StamopoulosPRB06}.
Indeed, very recently Monton et al. \cite{Monton07} studied Nb/Co
MLs in the parallel-field configuration and reported a behavior
similar to the one observed in
Refs.\onlinecite{StamopoulosPRB05,StamopoulosSST06,StamopoulosPRB06}.
These authors \cite{Monton07} interpreted their results based
exclusively on the influence of the FMs' stray fields on the SC
layers. However, that work \cite{Monton07} not only confirmed some
of the observations that were originally reported in
Refs.\onlinecite{StamopoulosPRB05,StamopoulosSST06,StamopoulosPRB06}
but expanded our knowledge to the complete mutual effect; the
authors \cite{Monton07} revealed that the screening coming from
the SC layers influences drastically the magnetic domain state of
the adjacent FM ones.

Referring to the normal-field configuration Moshchalkov and
colleagues revealed the importance of stray fields for the case of
Co-Pd/Nb/Co-Pd TLs, BaFe$_{12}$O$_{19}$-Nb BLs and
PbFe$_{12}$O$_{19}$-Nb BLs that were studied in
Refs.\onlinecite{Moshchalkov05}, \onlinecite{MoshchalkovNature}
and \onlinecite{MoshchalkovPRL}, respectively. In these works
\cite{Moshchalkov05,MoshchalkovNature,MoshchalkovPRL} where an
insulating interlayer was deposited between the SC and FM layers
in order the electronic coupling to be removed, the spatial
evolution of superconductivity nucleation above domain walls and
inside magnetic domains was studied extensively. Also, for simple
BLs we note that a magnetoresistance effect was observed by
Ryazanov et al. in Ref.\onlinecite{Ryazanov03} for the case of
Nb-Cu$_{0.43}$Ni$_{0.57}$ hybrids. These authors attributed the
magnetoresistance effect to the dissipation produced by the flow
of vortices since the SC's lower critical field is exceeded by the
FM's stray fields that emerge due to its multidomain magnetic
structure attained at coercivity. Very recently, in
Ref.\onlinecite{Aartsnew} Aarts and colleagues have concluded to
the same interpretation for the pronounced magnetoresistance
effect that they observed in all-amorphous GdNi-MoGe-GdNi TLs for
the parallel-field configuration.

Steiner and Ziemann in Ref.\onlinecite{Steiner06} have dealt with
the parallel-field configuration in a variety of hybrid TLs and
BLs consisting of Co, Fe and Nb by performing transport and {\it
longitudinal} magnetization measurements. These authors also
employed micromagnetic simulations to convincingly show that in
their samples the stray fields that emerge at coercivity are
related to the observed magnetoresistance peaks. However, the
magnetization measurements that were presented in
Ref.\onlinecite{Steiner06} referred only to the {\it longitudinal}
magnetic component and were limited in the {\it normal} state of
the SC constituent.

Experimental data on both the {\it longitudinal} and {\it
transverse} magnetic components that were obtained not only in the
{\it normal} but also in the {\it superconducting} state were
presented in our recent work, Ref.\onlinecite{StamopoulosRCPRBSub}
were we investigated these topics in relevant FM-SC-FM TLs and
FM-SC BLs that were constructed of low spin polarized
Ni$_{80}$Fe$_{20}$ and low-T$_c$ Nb. While the detailed results
that were presented in Ref.\onlinecite{StamopoulosRCPRBSub} refer
to plain TLs and BLs, the ones that were reported in
Ref.\onlinecite{StamopoulosPRLSub} relate to the respective
specimens that incorporate the mechanism of exchange bias. The
preliminary data that we reported in
Ref.\onlinecite{StamopoulosPRLSub} for exchange biased
Ni$_{80}$Fe$_{20}$-Nb-Ni$_{80}$Fe$_{20}$ TLs were in agreement
[contrast] to the ones presented in
Refs.\onlinecite{Pena05,Rusanov06,Visani07}
[Refs.\onlinecite{Gu02,Potenza05,Moraru06,Moraru06B}] and showed
evidence that the parallel [antiparallel] in-plane magnetization
configuration of the outer FM layers promotes [suppresses] the
transport properties of the SC interlayer. On the other hand, the
detailed results that were presented in
Ref.\onlinecite{StamopoulosRCPRBSub} for plain
Ni$_{80}$Fe$_{20}$-Nb-Ni$_{80}$Fe$_{20}$ TLs revealed a pronounced
dissipation peak that, although it was significantly stronger than
the one reported by Rusanov et al. in Ref.\onlinecite{Rusanov06}
for Ni$_{80}$Fe$_{20}$-Nb-Ni$_{80}$Fe$_{20}$ TLs of robust
uniaxial magnetic anisotropy, it was quantitatively identical to
the one reported by Pe\~{n}a et al., in Ref.\onlinecite{Pena05}
and Visani et al., in Ref.\onlinecite{Visani07} for the case of
La$_{0.7}$Ca$_{0.3}$MnO$_3$-YBa$_2$Cu$_3$O$_7$-La$_{0.7}$Ca$_{0.3}$MnO$_3$
TLs. However, in Ref.\onlinecite{StamopoulosRCPRBSub} we have
attributed the pronounced dissipation peaks to a different
underlying mechanism than the one reported in
Refs.\onlinecite{Pena05,Rusanov06,Visani07}. By performing
measurements on the TLs' {\it transverse} magnetic component in
both the normal and superconducting states, we clearly showed that
below its T$_c^{SC}$ the Nb interlayer under the influence of the
outer NiFe layers attains a magnetization component {\it
transverse} to the external magnetic field. When these {\it
transverse} magnetization data were compared to the transport ones
it was revealed that the pronounced magnetoresistance peaks that
are observed in the low-field regime are due to the suppression of
superconductivity by {\it transverse} stray fields that they
interconnect the outer Ni$_{80}$Fe$_{20}$ layers. Long time ago
relevant features were also observed in FM-IN-FM and FM-NM-FM TLs
(where IN and NM stand for insulator and non magnetic metal,
respectively). For instance, S. Parkin and colleagues
\cite{Gider98,Parkin00} have clearly shown that in such TLs of
IN/NM interlayer the magnetostatic interaction of the outer FM
layers through stray fields, that may be farther promoted by
surface roughness,\cite{Parkin00,Neel62,Demokritov94} could lead
to their significant overall magnetostatic coupling. This
stray-fields coupling plays a unique role in FM-IN-FM and FM-NM-FM
TLs since it alters drastically the distinct magnetic character of
the outer FM layers; ultimately this mechanism makes a
"magnetically hard" FM layer to rather behave as "magnetically
soft". Except for the influence of surface
roughness,\cite{Parkin00,Neel62,Demokritov94} the out-of-plane
rotation of the magnetization and the subsequent magnetostatic
coupling of the outer FM layers could be promoted or suppressed by
other parameters such as (a) the intrinsic magnetic anisotropy of
the specific FM materials, and (b) the extrinsic, shape-induced
magnetic anisotropy that is related to the thickness of the
employed FM layers. Returning to the case of our FM-SC-FM TLs we
stress that since the only difference between the FM-SC hybrids
studied in Refs.\onlinecite{StamopoulosRCPRBSub} and
\onlinecite{StamopoulosPRLSub} was the presence of exchange bias
it was natural to examine whether the mechanism that was employed
in Ref.\onlinecite{StamopoulosRCPRBSub} for the interpretation of
the experimental results obtained in plain TLs and BLs also holds
for the exchange biased ones that were only preliminary
highlighted in Ref.\onlinecite{StamopoulosPRLSub}.

This is the main topic of our current work: presenting detailed
magnetization and transport results that uncover the underlying
mechanism that motivates the relative promotion of
superconductivity in exchange biased FM-SC hybrids as it was
recently reported in Ref.\onlinecite{StamopoulosPRLSub}. We stress
that in this work we study the parallel-field configuration as was
done in all other
Refs.\onlinecite{Gu02,Potenza05,Moraru06,Moraru06B,Pena05,Rusanov06,Visani07,Steiner06,StamopoulosRCPRBSub,StamopoulosPRLSub,Rusanov05}
so that a straightforward comparison may be rightfully performed.
However, our proposals could be applicable to results that were
obtained for the normal-field configuration that was studied in
Refs.\onlinecite{Singh07,Moshchalkov05,MoshchalkovNature,MoshchalkovPRL,Ryazanov03}.
The presented magnetization data refer not only to the {\it
longitudinal} magnetic component (parallel to the externally
applied magnetic field) but also to the {\it transverse} one
(normal to the externally applied magnetic field). Furthermore, we
present data that are obtained not only in the {\it normal} regime
of the SC but also in its {\it superconducting} state. Such data
have not been presented in the literature and are shown here for
the first time. In agreement to the results obtained in
Ref.\onlinecite{StamopoulosRCPRBSub} for plain samples, here we
clearly show that even in exchange biased ones the {\it
transverse} magnetization data reveal crucial information for the
interpretation of the effects that are observed in the
magnetoresistance:

The extreme magnetoresistance peaks observed in the TLs are
primarily [secondarily] motivated by the magnetic coupling of the
outer FM layers as their magnetizations rotate out-of-plane
[in-plane] at the coercive field. The formed transverse
magnetization field that penetrates the SC interlayer ultimately
suppresses its superconducting properties since it primarily
[secondarily] exceeds its lower [upper] critical field. Referring
to the BLs although out-of-plane rotation of the magnetization of
the single FM layer is still observed, due to the absence of a
second FM layer in these structures magnetic coupling doesn't
occur so that the observed magnetoresistance peaks are only
modest. Consequently, in FM-SC-FM TLs the relative in-plane
magnetization configuration is not the only parameter that should
be examined, as was done in
Refs.\onlinecite{Gu02,Potenza05,Moraru06,Moraru06B,Pena05,Rusanov06,Visani07,StamopoulosPRLSub}.
{\it Primarily the out-of-plane magnetization configuration should
be considered as the underlying mechanism of the pronounced
magnetoresistance peaks.} This new experimental outcome assists us
toward understanding the contradictory transport data that have
been obtained recently in relevant FM-SC-FM
TLs.\cite{Gu02,Potenza05,Moraru06,Moraru06B,Pena05,Rusanov06,Visani07,StamopoulosPRLSub}

Most importantly, here we show that since the exchange bias
mechanism directly controls the in-plane magnetic order it also
controls the out-of-plane rotation of the FMs' magnetization so
that the magnetoresistance peaks may be tailored at will.
Uncontested evidence is presented through I-V characteristics for
the virgin and exchange biased states of the hybrids that exchange
bias relatively promotes the transport ability of the complete
hybrid. Also, the so-called training effect that accompanies the
exchange bias in the normal state is clearly preserved even in the
superconducting state and controls the transport behavior of the
SC.

Finally, a fair comparison of all the relevant experimental works
that have treated both exchange biased and plain FM-SC-FM TLs and
FM-SC BLs is carried out. Based on a prerequisite that should hold
for the coercive fields of the outer FM layers we propose a
plausible stray-fields mechanism that explains efficiently all the
available experiments that were brought in our attention in the
recent literature.

The present article is structured as following. Section II
presents the sample preparation techniques and experimental
details. Section III is devoted to the TL structures: subsection
III.A presents detailed magnetic characterization in order to
clearly reveal all the specific characteristics of the exchange
bias mechanism in our samples, subsections III.B and III.C present
comparatively transport and magnetization data for both the
longitudinal and transverse components, and in subsection III.D we
discuss the results obtained in our TLs and propose a stray-fields
coupling mechanism that considering it is based on a specific
requisite is also consistent with other current experiments.
Finally, in subsection III.E we show I-V characteristics for the
TLs. Section IV refers to BLs and is divided in subsection IV.A
which presents comparative transport and magnetization data, and
in subsection IV.B which is exclusively devoted to the training
effect that is observed in both the transport and magnetization
data. Finally, section V summarizes our observations and presents
the possible impact that these may have on current and future
experimental and theoretical works.

\section{Preparation of samples and experimental details}

The samples were sputtered on Si $[001]$ substrates under an Ar
environment ($99.999 \%$ pure). In order to eliminate the residual
oxygen that possibly existed in the chamber we performed Nb
pre-sputtering for very long times. \cite{Stamopoulos05PRB} During
the pre-sputtering process all the residual oxygen was absorbed by
the dummy Nb. This procedure has a direct impact on the quality of
the produced films.\cite{Stamopoulos05PRB} The Nb layers were
deposited by dc-sputtering at $46$ W and an Ar pressure of $3$
mTorr. More details may be found in
Ref.\onlinecite{Stamopoulos05PRB}. For the Ni$_{80}$Fe$_{20}$
(NiFe) layers rf-sputtering was employed at $30$ W and $4$ mTorr.
We should stress that: (i) all depositions were carried out {\it
at room temperature} and (ii) {\it no} magnetic field was applied
during the deposition of the NiFe layers. However, the samples
can't be shielded from the residual magnetic fields existing in
the chamber of our magnetically-assisted-sputtering unit.
Measurements by means of a Hall sensor revealed that at the place
where the substrates are mounted the residual fields exhibit
parallel components of magnitude $10-15$ Oe at maximum. Thus, our
NiFe films exhibit in-plane anisotropy. However, they don't
exhibit detectable uniaxial anisotropy since the magnetic field
sources are placed symmetrically on the perimeter of the circular
rf-gun (for more details see the magnetization results presented
in Ref.\onlinecite{StamopoulosRCPRBSub}). In order to inflict the
mechanism of exchange bias \cite{Bean57,Schuller99,Takano99} the
NiFe {\it bottom} layers were annealed systematically under O$_2$
($99.999 \%$ pure), Ar ($99.999 \%$ pure) or reductive Ar-H$_2$
($94 \%-6 \%$ mixture) atmosphere for durations in the range
$2-24$ hours at low temperatures $100-200$ C. Afterwards, the
complete BL or TL structures were deposited on top. As a result,
in the exchange biased structures the NiFe bottom layer possesses
exchange bias, while the top one (for the TLs) is always plain. In
the rest of this work we will call: exchange biased TLs as
BFM-SC-PFM, exchange biased BLs as BFM-SC, plain TLs as
PFM-SC-PFM, and finally plain BLs as PFM-SC ("B" stands for
"biased" and "P" for "plain"). The thicknesses are in the range
$20-40$ nm for the NiFe outer layers, while for Nb the thickness
varies between $40-60$ nm.

Our magnetoresistance measurements were performed by applying a
dc-transport current (always normal to the magnetic field) and
measuring the voltage in the standard four-point configuration. In
most of the measurements the applied current was $I_{{\rm
dc}}=0.5$ mA, which corresponds to an effective density $J_{{\rm
dc}}\approx 900$ A/cm$^2$ (typical in plane dimensions of the
films are $6\times 5$ mm$^2$). The temperature control and the
application of the magnetic fields were achieved in a
superconducting quantum interference device (SQUID) (Quantum
Design). In all cases the applied field was {\it parallel} to the
films.

\section{Exchange Biased TLs}

\subsection{Magnetic characterization of the samples}

\begin{figure}[tbp] \centering%
\includegraphics[angle=0,width=6.5cm]{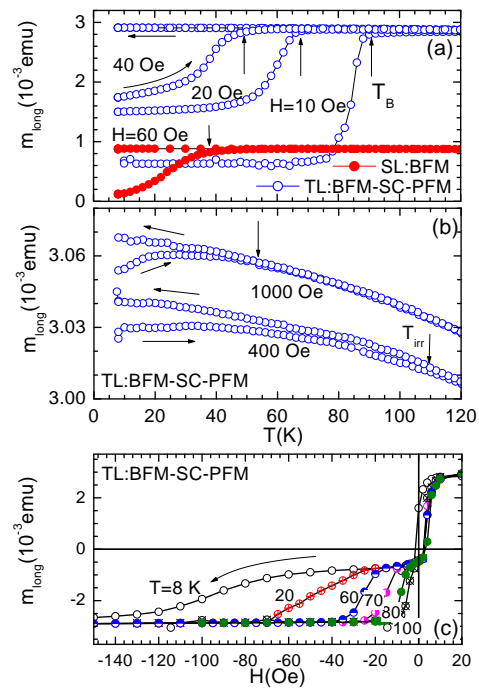}
\caption {(Colour online) Zero-field cooled and field cooled
isofield m(T) curves of the longitudinal component obtained in the
temperature range $8$ K$<$T$<120$ K for a TL:BFM-SC-PFM (open
circles) and a SL:BFM (solid circles) at relatively low (a) and
high (b) values of the external magnetic field. (c) Isothermal
loop measurements at various temperatures from above T$_B=90$ K
down to $8$ K. For the sake of clarity only the decreasing
branches are shown.}
\label{b1}%
\end{figure}%

The data presented in Figs. \ref{b1}(a)-\ref{b1}(c) refer to the
longitudinal magnetic component (parallel to the external magnetic
field which in turn is applied parallel to the specimen's surface
as it is schematically presented in the inset of Fig. \ref{b15},
below). Such data are needed for a thorough magnetic
characterization of the produced TLs. In panels (a) and (b) we
show both the zero-field cooled and field cooled isofield
m$_{long}$(T) curves obtained in the temperature range $8$
K$<$T$<120$ K for a TL at relatively low and high values of the
external magnetic field, respectively. In addition, we show one
representative curve for a NiFe BFM single layer (SL) (solid
circles) in order to stress that the behavior exhibited by the TL
is exclusively related to its bottom BFM NiFe layer. We clearly
see that a blocking temperature \cite{Schuller99,Takano99} may be
identified with T$_B=90$ K for this particular TL. In all the TLs
that have been studied the blocking temperature ranges between
$90$ K$<$T$_B<100$ K. As we see, above T$_B$ even the lowest
applied magnetic field saturates both the bottom and the top NiFe
layers, while below T$_B$ the BFM bottom layer exhibits a more
hard character and needs a few decades Oe to be saturated. As we
progressively apply higher magnetic fields the zero-field cooled
and field cooled curves merge; eventually at $1$ kOe only a minor
irreversibility exists which is difficult to be discerned (see the
emu range in panel (b)). The fact that the magnetic behavior of
the TL changes drastically below T$_B$ may also be revealed by
isothermal loop measurements as the ones presented in panel (c).
For the sake of clarity the presented data refer only to the
decreasing branch of the complete m(H) loops. We clearly see that
below T$_B=90$ K in addition to the reversal of the soft NiFe top
layer, which occurs around zero field, a second loop is
superimposed with a coercivity that gradually increases as we
lower the temperature. This particular second loop is related to
the BFM NiFe bottom layer.

\begin{figure}[tbp] \centering%
\includegraphics[angle=0,width=6.5cm]{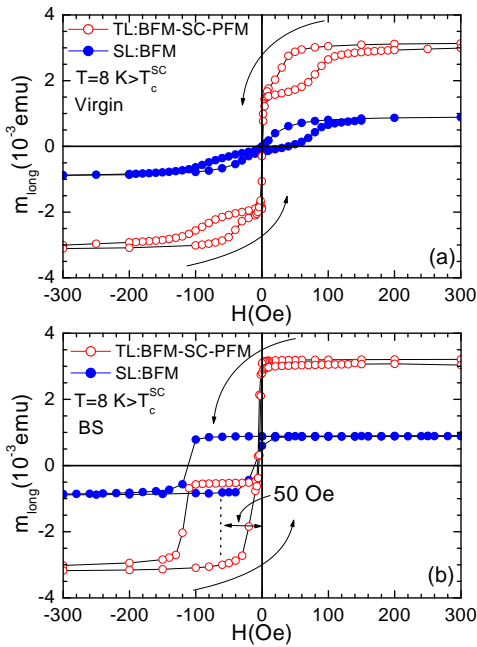}
\caption {(Colour online) Magnetization loops of the longitudinal
component obtained at T$=8$ K for a TL:BFM-SC-PFM (open circles)
and a SL:BFM (solid circles) when the samples were virgin (a) and
when exchange biased (b).}
\label{b2}%
\end{figure}%

We stress that although all data presented in Figs.
\ref{b1}(a)-\ref{b1}(c) show clear indications for the existence
of the exchange bias mechanism they refer to the case when the
specimens were virgin. However, the fingerprint of unidirectional
magnetic anisotropy is a shift observed in the magnetization loop
when the sample is exchange biased i.e. when it is cooled from
above the Neel or blocking temperature under the presence of a
sufficiently high magnetic field. \cite{Schuller99,Takano99} Such
data are presented in Figs. \ref{b2}(a) and \ref{b2}(b) for a
complete TL (open circles) and a SL (solid circles) at T$=8$
K$>$T$_c^{SC}$ when the samples were virgin and BS, respectively.
We see that in both cases the magnetization loop of the TL is the
superposition of the loop of the BFM bottom SL (shown) and of the
PFM top one (not shown). When the samples were biased by cooling
them from above T$_B=90$ K under the presence of H$_{ex}=1000$ Oe
the obtained loops are shifted toward negative fields by $50$ Oe.
We have to note that the shape of the loop of the BFM SL points
rather to a system comprised of hard and soft NiFe phases than an
one that constitutes of FM and AFM areas. However, as has been
revealed in recent years the underlying physics of exchange
coupled hard/soft and FM/AFM systems are essentially the same.
\cite{Fullerton98,Sort04,Mangin06,Fullerton99,Chien03} For
instance, Dieny and colleagues have shown very recently
\cite{Sort04} that in a soft NiFe layer which is exchange coupled
with a hard and strongly anisotropic Co/Pt multilayer all the
characteristics of exchange bias are met, including shifted
magnetization loops, training effects, decreasing bias field with
increasing temperature etc. Also, Chien et al.\cite{Chien03} have
revealed that in the archetypal NiFe/FeMn FM/AFM exchange bias
system the spins structure in the main body of the AFM layer is
not static but due to the coupling with the FM one the spins of
the FeMn layer form an exchange spring that may wind and unwind as
happens in typical soft/hard BLs. Finally, lately Ali et al. in
Ref.\onlinecite{Ali07} revealed that the effect of exchange bias
has even wider generality since they reported on the observation
of this effect in FM/spin glass BLs.

\begin{figure}[tbp] \centering%
\includegraphics[angle=0,width=6.5cm]{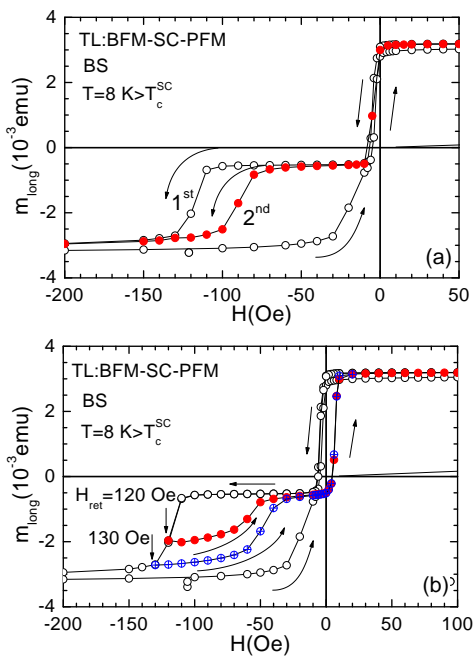}
\caption {(Colour online) (a) Complete and (b) minor successive
magnetization loops of the longitudinal component obtained at
T$=8$ K$>$T$_c^{SC}$ for a TL:BFM-SC-PFM.}
\label{b3}%
\end{figure}%

Although we have revealed the fingerprint of exchange bias in our
samples, in Figs. \ref{b3}(a) and \ref{b3}(b) we present
additional magnetization data necessary for a reliable magnetic
characterization. Panel (a) shows two successive loops obtained at
T$=8$ K$>$T$_c^{SC}$, while panel (b) presents minor successive
loops at the same temperature. The data presented in panel (a)
reveal the so-called training effect (see section $6.6$ in
Ref.\onlinecite{Schuller99}). The drastic reduction of the loop's
width observed in our data indicates that consecutive loops lead
to a softening of the underlying physical origin of exchange bias.
This result resembles strongly the one obtained by Gruyters and
Riegel \cite{Gruyters00} in a CoO/Co BL that is considered as a
very typical exchange bias system.\cite{Schuller99} This strong
similarity reveals that indeed the hard/soft and FM/AFM systems
have similar underlying physical principles;\cite{Fullerton99} all
the effects observed in both systems are motivated by the exchange
coupling occurring at the interfaces of the two different
constituents. This is further demonstrated by the data presented
in panel (b). While the first part of the loop where the soft NiFe
phase reverses its magnetization is reversible, the second part
clearly exhibits its irreversible nature associated to the
reversal of the hard phase magnetization.

\subsection{Comparison between transport and magnetization data}

\begin{figure}[tbp] \centering%
\includegraphics[angle=0,width=6.5cm]{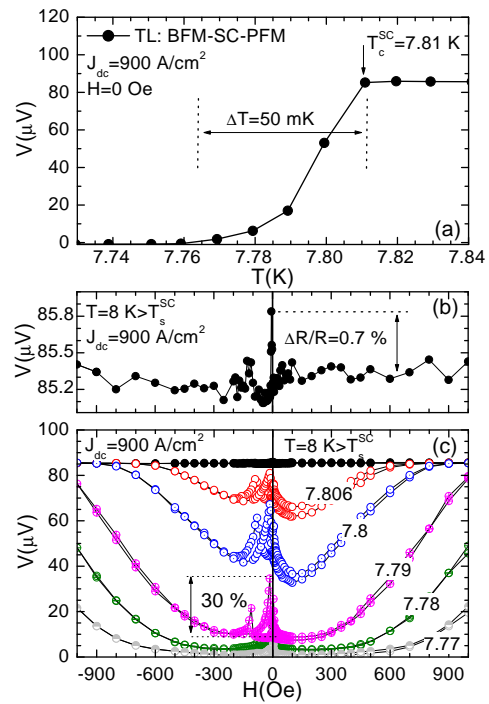}
\caption {(Colour online) (a) Zero-field resistive transition of a
TL:BFM-SC-PFM with T$_c^{SC}=7.81$ K. (b) Isothermal V(H) curve
obtained at T$=8$ K$>$T$_c^{SC}=7.81$ K. (c) Isothermal V(H)
curves obtained at temperatures allocated across the zero-field
isofield resistive V(T) curve presented in panel (a). All the data
were obtained when the TL was BS.}
\label{b4}%
\end{figure}%

Figures \ref{b4}(a)-\ref{b4}(c) present detailed transport data
for a TL:BFM-SC-PFM that were obtained very close to the
zero-field critical temperature T$_c^{SC}=7.81$ K of its SC
interlayer. Panel (a) shows the zero-field resistive transition of
the TL. The isothermal magnetoresistance V(H) curves presented in
panels (b) and (c) were obtained at temperatures allocated across
this specific zero-field resistive V(T) curve. Panel (b) presents
the obtained curve at T$=8$ K$>$T$_c^{SC}=7.81$ K. We see that
there is one distinct peak that is positioned at zero field and
another minor peak which occurs around $-120$ Oe. According to the
definition $(R_{max}-R_{min})/R_{nor}\times100\%$ that is commonly
used to categorize magnetoresistance effects
\cite{Pena05,Rusanov06,Visani07,StamopoulosRCPRBSub} the effect
observed in the normal state of our TL is very weak: $0.7\%$. In
contrast, when the Nb interlayer becomes SC the effect is strongly
enhanced as it is shown in panel (c). While as the applied field
is lowered the resistance of the TL is also reduced, for magnetic
fields in the range $-200$ Oe$<$H$<200$ Oe the measured resistance
presents pronounced peaks. These magnetoresistance peaks become
maximum at the middle of the zero-field resistive transition (see
panel (a)) and gradually diminish as both the normal and the fully
superconducting states are recovered. This result was reported
very recently in Ref.\onlinecite{StamopoulosPRLSub} for the same
BFM-SC-PFM TLs and reminisces the one reported in
Ref.\onlinecite{StamopoulosRCPRBSub} for more simple PFM-SC-PFM
TLs and the one reported in Refs.\onlinecite{Pena05} and
\onlinecite{Visani07} for
La$_{0.7}$Ca$_{0.3}$MnO$_3$-YBa$_2$Cu$_3$O$_7$-La$_{0.7}$Ca$_{0.3}$MnO$_3$
ones. The observed effect is strong. The percentage resistance
change $(R(0)-R(H))/R(H)\times100\%$ meets the quantitative
criteria for considered as a giant magnetoresistance (GMR) effect
\cite{Baibich88,Dieny94,Pena05,Visani07} since according to the
modest definition $(R_{max}-R_{min})/R_{nor}\times100\%$ (that
takes into account the reference resistance value obtained in the
normal state) it amounts to $50\%$. We note that this definition
should be considered as the most appropriate for the description
of the underlying physics since the alternative one
$(R_{max}-R_{min})/R_{min}\times100\%$ (that takes into account
the minimum value obtained in the superconducting state) suffers
from possible singularities originating from the zeroing of the
measured resistance that naturally occurs in every SC.

\begin{figure}[tbp] \centering%
\includegraphics[angle=0,width=6.5cm]{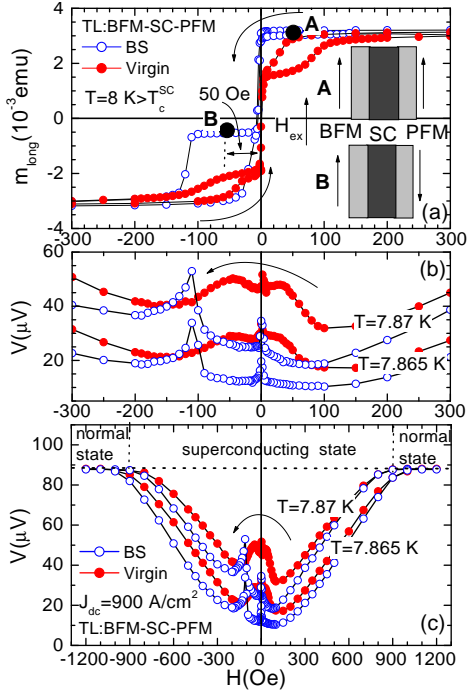}
\caption {(Colour online) (a) Detailed magnetization loops of the
longitudinal component for virgin (solid circles) and BS (open
circles) TL obtained at T$=8$ K$>$T$_c^{SC}=7.89$ K.
Representative magnetoresistance V(H) curves obtained just below
the critical temperature T$_c^{SC}=7.89$ K when the TL was virgin
(solid circles) and BS (open circles) in the low-field (b) and in
an extended field (c) regime. The insets show schematically the
relative in-plane magnetization configuration of the outer FM
layers at points A and B of the BS loop.}
\label{b5}%
\end{figure}%

We note that the V(H) curves presented in Figs.\ref{b4}(b) and
\ref{b4}(c) refer to the case when the TL was BS only {\it once},
prior to the first measurement. As we show below owing to the
training effect \cite{Schuller99} this measuring procedure leads
to significant differences when compared to the one when the
specimen gets always BS prior to each new measurement. In
addition, these curves differ strongly from the ones obtained for
virgin TL. This fact is clearly shown by the comparative data that
are presented in Figs.\ref{b5}(a)-\ref{b5}(c) around the
zero-field resistive transition of a second BFM-SC-PFM TL that
exhibits a slightly higher critical temperature, T$_c^{SC}=7.89$
K. For the sake of clarity panel (a) reproduces the detailed
magnetization loops for virgin (solid circles) and BS (open
circles) TL, while its insets show schematically the relative
in-plane magnetization configuration of the outer FM layers at
points A and B of the BS loop where H$_{ex}=50$ Oe and $-50$ Oe,
respectively. Panels (b) and (c) show representative
magnetoresistance isothermal data that were obtained at T$=7.865$
K and T$=7.87$ K in the low-field and in an extended field regime,
respectively. Both virgin (solid circles) and BS (open circles)
curves are shown for each temperature. Firstly, the virgin curve
mainly presents a peak around zero field that is significantly
broader than the one observed in PFM-SC-PFM TLs as it was reported
in Refs.\onlinecite{StamopoulosRCPRBSub}. An additional second
minor peak is also present at exactly zero field. As we see from
the virgin magnetization curve m$_{long}$(H) presented in panel
(a) the two outer NiFe layers reverse their magnetizations in the
respective field interval. This result was originally presented in
Refs.\onlinecite{StamopoulosPRLSub,StamopoulosRCPRBSub} and is
identical in magnitude to the one presented by Pe\~{n}a et al.
\cite{Pena05} and Visani et al. \cite{Visani07} for
La$_{0.7}$Ca$_{0.3}$MnO$_3$-YBa$_2$Cu$_3$O$_7$-La$_{0.7}$Ca$_{0.3}$MnO$_3$
TLs. {\it More importantly, here we clearly demonstrate that this
broad peak centered around zero field may be almost entirely
suppressed by the application of exchange bias.} Second, we see
that while in the normal state the virgin and BS curves clearly
coincide, as we enter in the superconducting state these curves
diverge, with {\it the BS curve placed significantly below the
virgin one} except for a small field regime around the coercive
field of the BS bottom NiFe layer (see panel (a)). In addition,
the BS magnetoresistance curve exhibits another peak placed at
zero field where the plain top NiFe layer reverses abruptly its
magnetization. Thus, we see that in both cases, virgin and BS, the
obtained magnetoresistance curves present two distinct peaks
placed at the field intervals where the two NiFe layers reverse
their magnetizations i.e. at the coercive fields. Consequently,
according to our results {\it the extreme dissipation peaks
observed in FM-SC-FM TLs are mainly related to the reversal of the
outer FM layers' magnetizations rather than their relative
in-plane magnetic configuration as it was supposed in
Refs.\onlinecite{Gu02,Potenza05,Moraru06,Moraru06B,Pena05,Rusanov06,Visani07}.}
Finally, the maximum value obtained in the BS magnetoresistance
curves is equal to the one obtained by the virgin ones. For
instance in Fig.\ref{b5}(b) we see that for T$=7.87$ K the maximum
value of the BS curve that is achieved at H$_{ex}=-110$ Oe equals
the one obtained by the respective virgin curve at H$_{ex}\approx
0$ Oe.

\begin{figure}[tbp] \centering%
\includegraphics[angle=0,width=6.5cm]{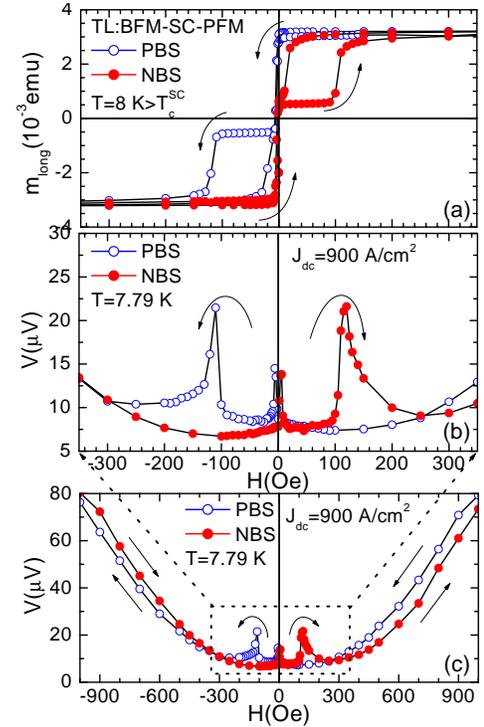}
\caption {(Colour online) (a) Magnetization loops of the
longitudinal component obtained at T$=8$ K$>$T$_c^{SC}$ when the
hybrid was positively BS (PBS) and when negatively BS (NBS). (b)
and (c) show the respective magnetoresistance V(H) curves obtained
at T$=7.79$ K$\approx$T$_c^{SC}$ in a focused and in an extended
field regime, respectively. In all cases the open circles refer to
PBS, while solid circles to NBS condition.}
\label{b7}%
\end{figure}%

In order to further clarify our results and to leave no doubt that
the effects observed in the superconducting state are motivated by
the exchange bias we performed additional test experiments. In
Figs.\ref{b7}(a)-\ref{b7}(c) we present data when the TL was
positively BS (PBS) and when negatively BS (NBS) under the
application of H$_{ex}=+1000$ Oe and H$_{ex}=-1000$ Oe,
respectively. Since the exchange bias imposes unidirectional
magnetic anisotropy the obtained curves should be exact mirror
images. This is what we have observed. Panel (a) shows the
magnetization loop m$_{long}$(H) curves of the longitudinal
magnetic component, while panels (b) and (c) show the isothermal
magnetoresistance V(H) curves that were obtained at T$=7.79$
K$\approx$ T$_c^{SC}$ in a narrow and in an extended field regime,
respectively. We clearly see that both the magnetization m(H) and
magnetoresistance V(H) curves that were obtained when the hybrid
is NBS are the exact mirror images of the ones obtained when the
TL was PBS. These results unexceptionably demonstrate that the
unidirectional magnetic anisotropy introduced by the exchange bias
dominates the behavior of the whole hybrid even in the
superconducting regime below T$_c^{SC}$.

\subsection{Magnetization data on the transverse component}

\begin{figure}[tbp] \centering%
\includegraphics[angle=0,width=6.5cm]{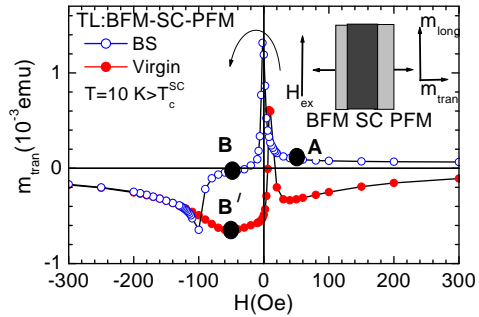}
\caption {(Colour online) Detailed magnetization data of the
transverse component for virgin (solid circles) and BS (open
circles) TL obtained at T$=10$ K$>$T$_c^{SC}$. The inset shows
schematically the experimental configuration between the external
field and the TL's magnetic components. Also, shown is the virgin
TL's specific out-of-plane magnetic configuration obtained around
zero field where the broad magnetoresistance peaks are observed
(see Figs.\ref{b5}(b)-\ref{b5}(c)).}
\label{b15}%
\end{figure}

For the case of PFM-SC-PFM TLs in
Ref.\onlinecite{StamopoulosRCPRBSub} we have showed that at the
same field intervals where the magnetoresistance peaks are
observed the magnetizations of the outer FM layers exhibit a
significant out-of-plane rotation. As a consequence the SC
interlayer below its T$_c^{SC}$ attains a m(H) model loop only in
its {\it transverse} magnetic component, while its {\it
longitudinal} one is almost completely suppressed. This proved
that for PFM-SC-PFM TLs the SC interlayer behaves diamagnetically
not in respect to the externally applied magnetic field but in
respect to the transverse magnetization field originating from the
magnetic coupling of the outer FM
layers.\cite{StamopoulosRCPRBSub} Examining this possibility for
the case of BFM-SC-PFM TLs could reveal important information
regarding the influence of exchange bias. In Fig.\ref{b15} we
present the respective data regarding the transverse magnetic
component of a BFM-SC-PFM TL when it was virgin (solid circles)
and when BS (open circles). The respective inset shows
schematically the experimental configuration between the external
field and the TL's magnetic components. Also, shown is the TL's
magnetic configuration around zero field for the magnetization
loop that is obtained when the TL is virgin. We recall that in
this field interval the broad magnetoresistance peaks are observed
for the virgin TL (see Figs.\ref{b5}(b)-\ref{b5}(c)). {\it
Interestingly, the respective transverse magnetization data
clearly show that in this field interval the out-of-plane magnetic
configuration of the outer FM layers is antiparallel.} By
comparing to the respective longitudinal magnetic data shown in
Fig.\ref{b5}(a) we clearly see that in both virgin and BS cases
the reversal of the outer FM layers' magnetization is accompanied
by a significant out-of-plane rotation occurring across the
coercive field. Most importantly, we see that the exchange bias
restricts the out-of-plane rotation to a narrow field range around
H$_{ex}=-100$ Oe, in contrast to the virgin case where its
out-of-plane rotation extends in the whole range of negative
fields $-200$ Oe$\leq$ H$_{ex}\leq 0$ Oe. The respective data of
the top PFM layer's magnetization don't show significant
differences.

\begin{figure}[tbp] \centering%
\includegraphics[angle=0,width=6.5cm]{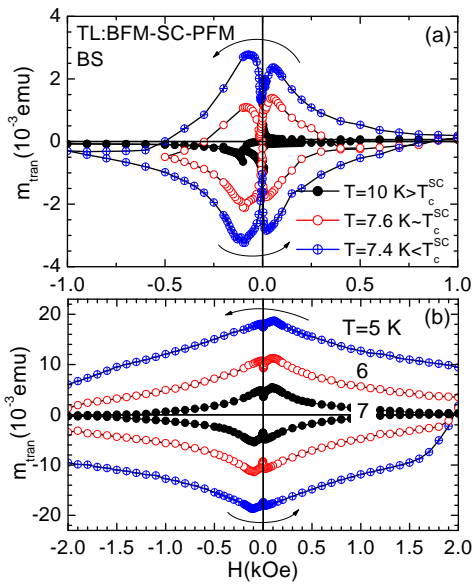}
\caption {(Colour online) Detailed magnetization loops of the BS
TL's transverse component obtained (a) close to T$_c^{SC}$ and (b)
well inside the superconducting state.}
\label{b16}%
\end{figure}

Based on the results shown in Fig.\ref{b15} we may assume that the
dissipation peaks presented in Figs.\ref{b5}(b)-\ref{b5}(c) could
be related to the out-of-plane rotation. However, the measurements
presented in Fig.\ref{b15} refer to the {\it normal} state of the
SC interlayer since they were obtained at T$=10$ K$>$T$_c^{SC}$.
To further examine the effect we performed detailed measurements
of the TL's transverse magnetic component in the {\it
superconducting} regime. Complete data are presented in
Figs.\ref{b16}(a)-\ref{b16}(b) for temperatures close [upper
panel] and well below [lower panel] T$_c^{SC}$. We clearly see
that the TL's transverse magnetic component exhibits the model
loop expected for a SC when bulk pinning dominates. Also, we see
that in the low-field regime a suppression of the magnetization is
observed owing to the out-of-plane rotation of the FM layers'
magnetizations.

\begin{figure}[tbp] \centering%
\includegraphics[angle=0,width=6.5cm]{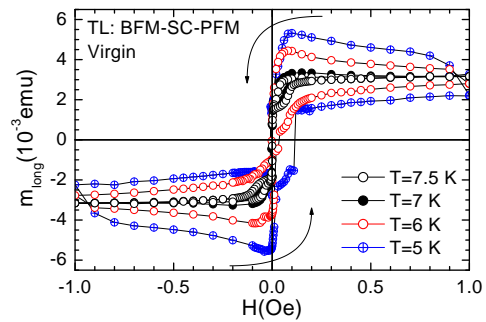}
\caption {(Colour online) Detailed magnetization loops of the BS
TL's longitudinal component obtained close to T$_c^{SC}$ and well
inside the superconducting state.}
\label{b20}%
\end{figure}

In contrast, the measured loop of the longitudinal magnetic
component in the superconducting state is almost identical to the
one observed in the normal regime. The respective data for the
longitudinal component obtained at temperatures close and well
below T$_c^{SC}$ are shown in Fig.\ref{b20}. We see that the
magnetization loop of the complete TL qualitatively reminisces not
the one of a SC but that of a FM. It is only well below T$^c_{SC}$
where significant hysteresis shows up in the high field regime.
However, close to zero field the typical FM behavior is recovered
in all cases. The results presented in
Figs.\ref{b16}(a)-\ref{b16}(b) and \ref{b20} clearly demonstrate
that {\it  the SC behaves diamagnetically not in respect to the
externally applied magnetic field but in respect to a new
transverse magnetic field that obviously emerges due to the
magnetic coupling of the outer FM layers (see the discussion
below).}

\begin{figure}[tbp] \centering%
\includegraphics[angle=0,width=6.5cm]{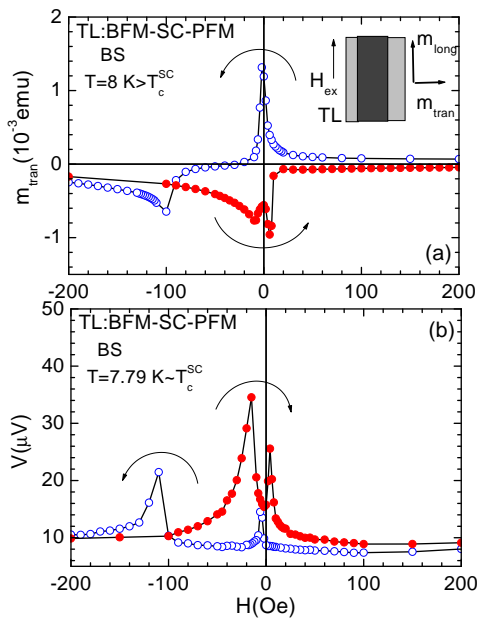}
\caption {(Colour online) Comparative presentation of the low
field transverse magnetic component (a) and magnetoresistance
curve (b) for a BS TL obtained at T$=8$ K$>$T$_c^{SC}$ and at
T$=7.79\approx$T$_c^{SC}$, respectively. Both increasing (solid
circles) and decreasing (open circles) branches are shown. The
inset shows schematically the experimental configuration between
the external field and the TL's magnetic components.}
\label{b17}%
\end{figure}

Figures \ref{b17}(a)-\ref{b17}(b) present comparatively low-field
transverse magnetization and transport data for a BFM-SC-PFM TL
when it is BS for both increasing (solid circles) and decreasing
(open circles) the external magnetic field. We clearly see that
the morphology of the presented curves is identical: the
dissipation peaks occur at exactly the same field values where the
transverse component gets maximum. This shows unambiguously that
{\it the dissipation peaks are related to the out-of-plane
rotation of the outer FM layers' magnetization}. Consequently, as
in the PFM-SC-PFM TLs studied in
Ref.\onlinecite{StamopoulosRCPRBSub} even for the BFM-SC-PFM TLs
studied in this work the pronounced dissipation peaks may be
ascribed to the strict influence of the transverse magnetization
field that emerge due to the magnetic interconnection of the outer
FM layers.

Ultimately, the magnetostatic coupling discussed here could
motivate the magnetoresistance peaks that we observe as following:
the stray fields that interconnect the outer NiFe layers penetrate
completely the Nb interlayer (see Fig.\ref{b22}, below). Depending
on the specific characteristics of the employed FM layers
(exchange field, size of magnetic domains, width and kind -Neel or
Bloch- of domain walls, etc) and of the SC interlayer (lower and
upper critical field values, bulk pinning force, etc) the FMs'
stray fields will primarily exceed the SC's lower critical field
and secondarily, in case that they are intense, even the SC's
upper critical field could {\it locally} be exceeded, at least
extremely close to its T$_c^{SC}$ where it attains low values. In
case where only the SC's lower critical field is exceeded by the
FMs' stray fields dissipation should set in due to flow of
vortices that are spontaneously created. As we lower the
temperature the magnetoresistance peaks become progressively
smaller and eventually disappear since either strong bulk pinning
emerges so that vortices are no longer free to move, or the SC's
increasing lower critical field exceeds the FMs' stray fields so
that a dissipationless Meissner state is ultimately recovered.
This explanation resembles the one that was presented by Ryazanov
et al. in Ref.\onlinecite{Ryazanov03} for
Nb-Cu$_{0.43}$Ni$_{0.57}$ BLs, and by Bell et al. in
Ref.\onlinecite{Aartsnew} for all-amorphous GdNi-MoGe-GdNi TLs.
Thus, our results extend the ones of Ref.\onlinecite{Aartsnew} to
non-amorphous NiFe-Nb-NiFe TLs. In case where also the SC's upper
critical field is exceeded the {\it localized} normal areas that
emerge should contribute extra dissipation. Eventually, in this
scenario the disappearance of the magnetoresistance peaks is owing
to the fact that as we progressively lower the temperature the
SC's upper critical field exceeds the stray fields that
interconnect the outer FMs so that bulk superconductivity is
completely restored throughout the whole SC interlayer. Of course,
since the SC's lower critical field is always smaller than its
upper critical one we expect that the mechanism described in the
former [latter] case should have a major [minor] contribution to
the observed magnetoresistance effect.

\subsection{FM-SC-FM TLs: comparison with current experiments}

As already discussed in the "Introduction" specific discrepancies
have been reported in the recent literature
\cite{Gu02,Potenza05,Moraru06,Moraru06B,Pena05,Rusanov06,Visani07,StamopoulosPRLSub}
regarding the operation of relevant FM-SC-FM TLs. Such TLs have
been considered as superconducting "spin valves" both
theoretically (see Refs.\onlinecite{Buzdin99,Tagirov99}) and
experimentally (see
Refs.\onlinecite{Gu02,Potenza05,Moraru06,Moraru06B,Pena05,Rusanov06,Visani07}).
We stress that we use the quotation marks since according to the
results of this work the physical mechanism motivating the
pronounced magnetoresistance peaks is not actually related to a
spin-dependent pair-breaking effect as it was theoretically
proposed in Refs.\onlinecite{Buzdin99,Tagirov99}. According to our
results the magnetoresistance peaks originate mainly from the
suppression of the SC interlayer's properties owing to the stray
fields that interconnect the outer FM layers, or in other words
owing to their out-of-plane magnetic configuration. Thus, the
experiments of
Refs.\onlinecite{Gu02,Potenza05,Moraru06,Moraru06B,Pena05,Rusanov06,Visani07,StamopoulosPRLSub}
should be reconsidered as being realizations of the theoretical
proposals presented in Refs.\onlinecite{Buzdin99,Tagirov99}. In
the discussion that follows we temporarily neglect the new
experimental results presented in this work regarding the
influence of the FMs' out-of-plane magnetization field on the
properties of the SC interlayer in order to make a brief overview
of the recent experimental works (see
Refs.\onlinecite{Gu02,Potenza05,Moraru06,Moraru06B,Pena05,Rusanov06,Visani07,StamopoulosPRLSub})
and to ultimately suggest an explanation that is consistent with
both the already available data (see
Refs.\onlinecite{Gu02,Potenza05,Moraru06,Moraru06B,Pena05,Rusanov06,Visani07,StamopoulosPRLSub})
and the new results that we present. For the moment we will assume
that only the in-plane relative magnetization configuration of the
outer FM layers controls the properties of the SC interlayer and
we will adopt the point of view proposed in
Refs.\onlinecite{Gu02,Potenza05,Moraru06,Moraru06B,Pena05,Rusanov06,Visani07,StamopoulosPRLSub}.

\subsubsection{Considering only the in-plane relative magnetization configuration}

\begin{figure}[tbp] \centering%
\includegraphics[angle=0,width=6.5cm]{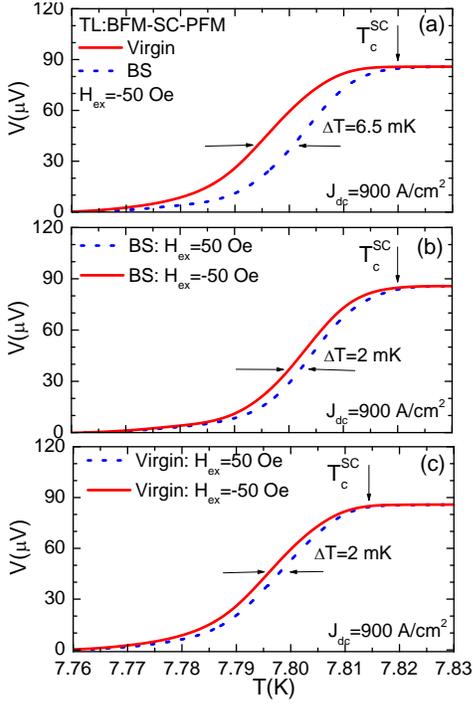}
\caption {(Colour online) Representative voltage curves as a
function of temperature (a) at $H_{ex}=-50$ Oe when the TL is
virgin (solid line) and BS (dotted line), (b) at $H_{ex}=-50$ Oe
(solid line) and $50$ Oe (dotted line) when it is BS, and (c) at
$H_{ex}=-50$ Oe (solid line) and $50$ Oe (dotted line) when it is
virgin.}
\label{b6}%
\end{figure}

While some works argued that the antiparallel [parallel] in-plane
magnetization configuration of the outer FM layers promotes
[suppresses] superconductivity (see
Refs.\onlinecite{Gu02,Potenza05,Moraru06,Moraru06B}), others have
provided evidence for exactly the opposite behavior where the
antiparallel [parallel] in-plane magnetization configuration of
the outer FM layers suppresses [promotes] superconductivity (see
Refs.\onlinecite{Pena05,Rusanov06,Visani07,StamopoulosPRLSub}). In
order to investigate this discrepancy we performed additional
experiments. Representative results are shown in
Figs.\ref{b6}(a)-\ref{b6}(c). First, these data reveal that {\it
the exchange bias clearly promotes the resistive critical
temperature of the TL} as for instance may be seen in panel (a)
where both V(T) curves were obtained for the same external field
H$_{ex}=-50$ Oe. The data shown in panels (b) and (c) were
obtained for opposite external fields H$_{ex}=-50$ Oe and
H$_{ex}=50$ Oe and refer to BS and virgin TL, respectively. The
two cases of the relative in-plane magnetization configurations of
the outer FM layers are realized: in both panels (b) and (c) the
antiparallel [solid lines] occurs for $H_{ex}=-50$ Oe and the
parallel [dotted lines] is obtained for $H_{ex}=50$ Oe. Of course,
the antiparallel configuration is more robust for the BS case as
it is presented in points A and B that refer to the BS loop shown
in Fig.\ref{b5}(a). We see that whether the bottom NiFe layer is
BS [panel (b)] or virgin [panel (c)] the antiparallel
[$H_{ex}=-50$ Oe] in-plane magnetization configuration of the
outer FM layers places the resistive transition at lower
temperatures compared to the case when the NiFe layers are
parallel [$H_{ex}=50$ Oe]. The same results were obtained for all
field values in the interval $-100$ Oe$<$H$_{ex}<100$ Oe, where
the antiparallel and parallel in-plane magnetization configuration
is realizable. Finally, we have to stress that the data presented
in Figs.\ref{b6}(a)-\ref{b6}(c) clearly prove that {\it the
observed effects are an exclusive property of the superconducting
state since in the normal state, i.e. for T$>$T$_c^{SC}$, the
resistance curves clearly coincide.}

\begin{figure}[tbp] \centering%
\includegraphics[angle=0,width=6.5cm]{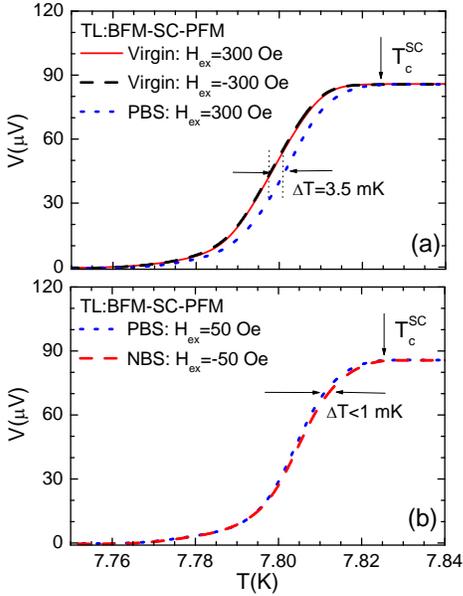}
\caption {(Colour online) (a) Magnetoresistance curves obtained at
the symmetric points $H_{ex}=-300$ Oe (dashed line) and $300$ Oe
(solid line) of the virgin TL's magnetization loop (see the virgin
loop in Fig.\ref{b5}(a)) and the respective curve obtained at
$H_{ex}=300$ Oe (dotted line) when the TL was positively BS. (b)
Test magnetoresistance data obtained for $H_{ex}=50$ Oe and $-50$
Oe when the TL was positively (dotted line) and negatively (dashed
line) BS (see the respective loops in Fig.\ref{b7}(a)).}
\label{b8}%
\end{figure}

Since in the measurements presented in
Figs.\ref{b6}(a)-\ref{b6}(c) the observed temperature shifts are
very small but, at least, comparable to the ones presented in
Refs.\onlinecite{Gu02,Potenza05,Moraru06,Moraru06B,Pena05,Rusanov06}
we performed some test measurements in order to ensure that these
shifts are motivated by the physics of the studied systems and are
not artifacts. The results of Figs.\ref{b8}(a)-\ref{b8}(b) serve
this aim. Panel (a) clearly shows that when the TL is virgin, at
the symmetric points $H_{ex}=-300$ Oe (dashed line) and $300$ Oe
(solid line) of the magnetization loop (see the virgin loop in
Fig.\ref{b5}(a)) the obtained V(T) curves clearly coincide as they
should. In addition, in the same panel presented is the respective
curve obtained at $H_{ex}=300$ Oe (dotted line) when the TL was
positively BS (PBS), i.e. under the application of a positive
magnetic field. Clearly the PBS resistive curve is placed at
higher temperatures compared to the virgin ones. Panel (b) shows
test data obtained for $H_{ex}=50$ Oe and $-50$ Oe when the TL was
PBS (positively BS) and NBS (negatively BS). The respective m(H)
loops are presented in Fig.\ref{b7}(a). Since the observed
divergence of the two curves is below $1$ mK we may assume that
the obtained V(T) curves coincide as they should.

\begin{figure}[tbp] \centering%
\includegraphics[angle=0,width=6.5cm]{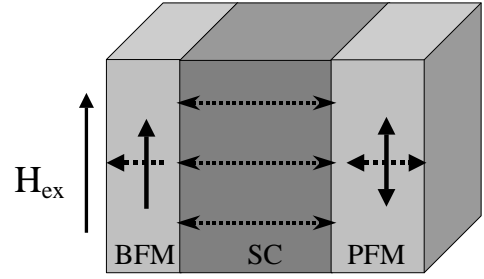}
\caption {(Colour online) Schematic representation of how the
outer FM layers may influence the SC interlayer. In most
experiments the BFM layer that is employed has a robust
magnetization due to pinning by exchange bias. In contrast, the
magnetization of the PFM layer is free to rotate. In our samples
both the in-plane (solid arrows) and out-of-plane (dotted arrows)
relative magnetization configurations of the outer FM layers are
present and influence the transport properties of the SC
interlayer.}
\label{b22}%
\end{figure}

According to the point of view adopted in
Refs.\onlinecite{Gu02,Potenza05,Moraru06,Moraru06B,Pena05,Rusanov06,Visani07,StamopoulosPRLSub}
where only the in-plane relative magnetization configuration of
the FMs was considered our results support the experimental
outcome of
Refs.\onlinecite{Pena05,Rusanov06,Visani07,StamopoulosPRLSub}
where it was concluded that the antiparallel [parallel] in-plane
magnetization configuration of the outer FM layers suppresses
[promotes] superconductivity. However, in this work we clearly
show that the underlying mechanism motivating the
magnetoresistance peaks is surely related to the {\it
out-of-plane} rotation of the magnetizations of the outer FM
layers and not only to their {\it in-plane} relative
configuration. The combination of both the in-plane and
out-of-plane relative magnetization configuration is schematically
presented in Fig.\ref{b22}. Thus, it is now easy to explain all
the available experimental data presented in
Refs.\onlinecite{Gu02,Potenza05,Moraru06,Moraru06B,Pena05,Rusanov06,Visani07}
and in this work.

\subsubsection{Considering both the in-plane and out-of-plane relative magnetization configuration}

First, let us consider the results presented in Fig.\ref{b6}(a) in
comparison to the ones of Fig.\ref{b15}. In Fig.\ref{b6}(a) both
curves were obtained for $H_{ex}=-50$ Oe but the BS curve is
placed significantly above the virgin one; $\Delta T=6.5$ mK.
Figure \ref{b15} reveals that at $H_{ex}=-50$ Oe when the TL is
virgin a significant part of its magnetization rotates
out-of-plane, while when it is BS the in-plane magnetic order is
preserved so that its transverse component is zero (see points
B$^{/}$ and B in Fig.\ref{b15}, respectively). Thus, the
transverse magnetic component of the outer FMs could strongly
influence the SC interlayer only for the virgin case, while for
the BS one its influence (if any) should be minimum.

On the other hand the data presented in Fig.\ref{b6}(b) refer to
the case when although the TL is always BS the external field's
polarity is opposite ($H_{ex}=50$ Oe vs $-50$ Oe). According to
the results of Fig.\ref{b15} for these two cases the difference of
the out-of-plane magnetic component is only minor (see points A
and B in Fig.\ref{b15}). This is the reason why the initial
temperature shift of $\Delta T=6.5$ mK between the curves
presented in Fig.\ref{b6}(a) decreases down to $\Delta T=2$ mK for
those presented in Fig.\ref{b6}(b). By making the assumption that
in both cases presented in Fig.\ref{b6}(b) the transverse
component is almost identical (see points A and B in
Fig.\ref{b15}) we may attribute the observed shift of $\Delta T=2$
mK exclusively to the in-plane relative magnetization
configuration of the outer FM layers. Under this assumption our
results are compatible to the ones presented in
Refs.\onlinecite{Pena05,Rusanov06,Visani07} and
\onlinecite{StamopoulosPRLSub} where it was concluded that the
antiparallel [parallel] in-plane magnetization configuration of
the outer FM layers suppresses [promotes] superconductivity.
However, as we demonstrate here small temperature shifts ($\Delta
T=2$ mK) of the resistive curves which are motivated mainly by the
relative in-plane magnetization configuration of the outer FM
layers are enhanced ($\Delta T=6.5$ mK) under the influence of the
out-of-plane rotation of their magnetizations.

\subsubsection{Prerequisite for intense stray-fields magnetoresistance in FM-SC-FM TLs}

The results that were presented until now and the discussion made
just above prove that since three parameters do get involved in
such hybrid TLs: (i) the out-of-plane magnetizations' rotation and
the unavoidable stray-fields coupling of the outer FM layers, (ii)
the in-plane relative magnetization configuration of the outer FM
layers, and (iii) the direct or indirect influence of exchange
bias, the small temperature shifts that are observed should not be
ascribed exclusively to the one of them, namely the in-plane
relative magnetic configuration, as it was suggested in
Refs.\onlinecite{Gu02,Potenza05,Moraru06,Moraru06B,Pena05,Rusanov06,Visani07}
and \onlinecite{StamopoulosPRLSub}.

Since we reveal the significance of the out-of-plane magnetic
component that is related to the stray fields emerging at magnetic
domain walls around coercivity, in this subsection let us first
make a brief overview of the literature related with this topic.
Subsequently, we propose a plausible explanation that in a
phenomenological level explains naturally all the recent
experimental works.

{\it Parallel field configuration:} First we discuss the
configuration where the external field is applied {\it parallel}
to a FM-SC hybrid as in the present work. The influence of stray
fields that originate from a FM multilayer on an adjacent SC was
already highlighted in
Refs.\onlinecite{StamopoulosPRB05,StamopoulosSST06,StamopoulosPRB06}.
Also, the influence of stray fields in FM-SC MLs was reported very
recently in Ref.\onlinecite{Monton07}. However, for more simple
structures as the FM-SC-FM TLs studied in the present article the
dominant influence of stray-fields coupling of the outer FM layers
on the SC interlayer has been proven very recently in
Refs.\onlinecite{Steiner06,StamopoulosRCPRBSub} for the case of
FM-SC-FM TLs when the external magnetic field was applied {\it
parallel} to their surface. Detailed discussion on the possible
underlying mechanism that eventually suppresses the transport
properties of the "magnetically pierced" SC interlayer was made
above (see subsection III.C).

{\it Normal field configuration:} On the other hand the case of
FM-SC-FM TLs where the magnetic field was applied {\it normal} to
their surface it was studied in Refs.\onlinecite{Singh07} and
\onlinecite{Moshchalkov05} for Co-Pt/Nb/Co-Pt and Co-Pd/Nb/Co-Pd,
respectively. Unfortunately, in these works
\cite{Singh07,Moshchalkov05} complete transport data that would
reveal the behavior of the TL as the magnetic field is scanned
from positive to negative saturation of the FM layers were not
presented. Thus, the exact transport behavior around the FM's
coercive fields is still not clear for the normal-field
configuration. More specifically, Singh et al. in
Ref.\onlinecite{Singh07} contrasted their results with the ones
presented by Gillijns et al. in Ref.\onlinecite{Moshchalkov05},
and claimed that stray fields were not involved in their
experiments as this was evidenced when an insulating SiO$_2$ layer
was introduced at each FM-SC interface so that the electronic
coupling of the two constituents was removed. However, in the
specific transport experiments that were presented in
Ref.\onlinecite{Singh07} each FM layer was in its saturated
magnetic state so that possible stray fields were only limited at
the specimen's edges. This specific magnetic realization was
employed successfully long time ago in Ref.\onlinecite{ClintonAPL}
as the critical current of a SC strip was controlled by the stray
fields that were restricted at the edges of a homogeneously
magnetized FM layer that was adjacent to the SC through a thin
oxide interlayer.

We have to stress that in our work we refer not to the stray
fields that exist {\it only near the edges} of a homogeneously
magnetized FM as it was discussed in
Refs.\onlinecite{Singh07,ClintonAPL} but to the stray fields that
emerge naturally {\it all over the surface} of the FM owing to the
appearance of magnetic domains around its coercivity. This is a
very important difference that should not be disregarded. Here,
based on this stray-fields concept we propose a plausible
explanation that should hold for both the parallel and normal
field configurations and could efficiently capture all the
available experimental data presented in the recent literature
(see
Refs.\onlinecite{Gu02,Potenza05,Moraru06,Moraru06B,Pena05,Rusanov06,Visani07}
and \onlinecite{Steiner06,StamopoulosRCPRBSub,StamopoulosPRLSub}).

Intuitively, in a FM-SC-FM TL the magnetic coupling of the outer
FM layers mediated by the stray fields that emerge naturally near
coercivity should be maximum when the coercive fields of the FMs
are almost equal.\cite{Stamopoulosnewsub} When this condition is
fulfilled the two FM layers are susceptible to get magnetically
coupled by the stray fields accompanying the magnetic domains that
emerge {\it simultaneously} all over the surface of both FM layers
around their common coercive field. This proposition explains
naturally the occurrence of pronounced magnetoresistance peaks in
Refs.\onlinecite{Pena05,Visani07,Steiner06,StamopoulosRCPRBSub}.
In all these works the FM layers participating a TL had almost
identical coercive fields as this is evidenced by the only minor
two-step behavior observed in the presented m(H) loops. In
contrast, when the coercive field of the first FM layer is
significantly different than that of the second one the
stray-fields magnetic coupling is not intense so that the
magnetoresistance effect is getting weak. This is evidenced by the
experimental results presented in
Refs.\onlinecite{Gu02,Potenza05,Moraru06,Moraru06B,Rusanov06}. In
Ref.\onlinecite{Rusanov06} the two FM layers exhibit a few tens Oe
different coercive fields as this is clear by the distinct
two-step behavior in the presented m(H) loops. Furthermore, in
Refs.\onlinecite{Gu02,Potenza05,Moraru06,Moraru06B} the coercive
fields of the outer FM layers differ by several hundreds Oe.
Summarizing these data, compared to the magnetoresistance effect
reported in
Refs.\onlinecite{Pena05,Visani07,Steiner06,StamopoulosRCPRBSub},
where the FM layers share almost common coercive fields, the one
reported in Ref.\onlinecite{Rusanov06}, where distinct coercive
fields of the FMs are clearly resolved, is weaker. Eventually, in
Refs.\onlinecite{Gu02,Potenza05,Moraru06,Moraru06B}, where the
coercive fields are very different, notable magnetoresistance
peaks were not reported. Thus, it is natural to assume that when
the coercive regimes of the two FM layers exhibit significant
overlapping the observed magnetoresistance effect is pronounced,
while as the coercivities get progressively different the
magnetoresistance peaks eventually disappear. We believe that the
only mechanism that could be invoked for the consistent
interpretation of all these experimental data that are reported in
the recent literature
\cite{Gu02,Potenza05,Moraru06,Moraru06B,Pena05,Rusanov06,Visani07,Steiner06,StamopoulosRCPRBSub,StamopoulosPRLSub}
is the stray-fields coupling of the outer FM layers that "pierces"
magnetically the SC interlayer as it was discussed in
Ref.\onlinecite{StamopoulosRCPRBSub} for plain TLs and in this
work for exchange biased ones (see subsection III.C).

Now, the question that waits for an answer is what is the exact
role of the exchange bias mechanism in the BFM-SC-PFM TLs that are
studied in our work for the parallel field configuration? We
believe that the answer relies on the physical origin of exchange
bias and could be given by simple phenomenological arguments.
Since the exchange bias induces an in-plane unidirectional
anisotropy it also controls the magnetic in-plane order of the
TL's bottom BFM layer.\cite{Schuller99} Thus, when the bottom
layer is BS it maintains an in-plane magnetization for a wider
field range so that it doesn't exhibit significant out-of-plane
rotation, an ingredient that is necessary for the occurrence of
broad dissipation peaks. In this case the dissipation peaks are
restricted in a narrow field regime around the coercive field. In
contrast, when the bottom BFM layer is not exchange biased it is
free to attain a significant out-of-plane component. Thus, its
magnetic interconnection with the other PFM layer may be achieved
easily and the broad dissipation peaks evolve. Finally, as it was
proposed in Ref.\onlinecite{Stamopoulosnewsub} and discussed right
above the stray-fields coupling of the outer FM layers should be
more efficient under a specific prerequisite: when the FMs share
almost common coercive fields. Under this sense {\it the exchange
bias may be considered as a parameter that tunes the coercive
field of the one FM layer so that the stray-fields coupling may be
controlled through the realization of this specific prerequisite}.

We stress that in even more simple FM-IN-FM and FM-NM-FM TLs (IN
and NM stand for insulator and non magnetic metal, respectively)
both in-plane and out-of-plane magnetic mechanisms are getting
involved. For instance, for the parallel field configuration in
FM-IN-FM and FM-NM-FM TLs magnetostatic interactions of the outer
FM layers through stray fields that occur at domain walls could
lead to their significant coupling. Indeed, S. Parkin and
colleagues have shown \cite{Gider98,Parkin00} that such
stray-fields coupling occurring at domain walls plays a unique
role in FM-IN-FM and FM-NM-FM TLs since it vitiates potently the
distinct magnetic character of even quite different outer FM
layers. In addition, the action of stray fields could also be
promoted by roughness that probably exists at the interfaces
(whether it is correlated or
not).\cite{StamopoulosRCPRBSub,Neel62,Demokritov94}

\subsection{Dynamic behavior through I-V characteristics}

\begin{figure}[tbp] \centering%
\includegraphics[angle=0,width=6.5cm]{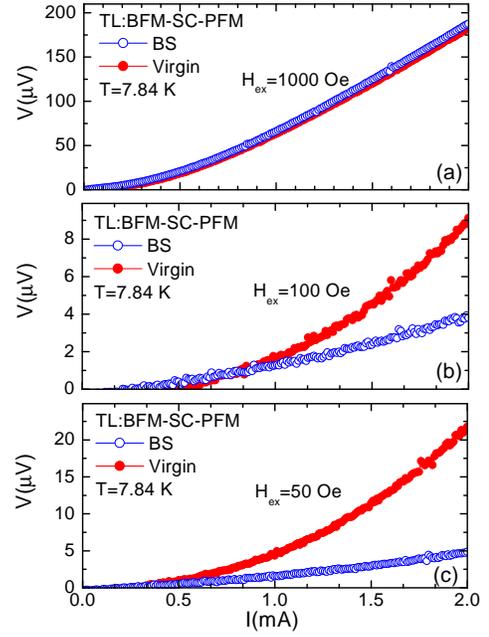}
\caption {(Colour online) I-V characteristics obtained at
temperature T$=7.84$ K and for magnetic fields (a) H$_{ex}=1000$,
(b) $100$ and (c) $50$ Oe. Data that were obtained for both virgin
(solid circles) and BS (open circles) bottom NiFe layer are
presented.}
\label{b9}%
\end{figure}%

In all the data presented so far we have shown that the exchange
bias promotes the superconducting properties of the complete
hybrid when compared to the case where the bottom NiFe layer is
virgin. Important information for the dynamic transport behavior
of our hybrids may be gained by measuring I-V characteristics.
Representative data are shown in Figs.\ref{b9}(a)-\ref{b9}(c) for
the specific temperature T$=7.84$ K and for magnetic fields
H$_{ex}=1000, 100$ and $50$ Oe. We clearly see that close to the
normal state i.e. for H$_{ex}=1000$ Oe there is no actual
distinction between the virgin (solid circles) and BS (open
circles) I-V curves. But as the applied field is lowered i.e. as
we enter deeper in the superconducting state we clearly see that
the BS I-V characteristics are placed much lower than the virgin
ones. This result reveals that the current-carrying capability of
the complete hybrid is strongly enhanced when the bottom NiFe
layer is BS compared to when it is virgin.

\section{Exchange Biased BLs}

\subsection{Comparison between transport and magnetization data}

\begin{figure}[tbp] \centering%
\includegraphics[angle=0,width=6.5cm]{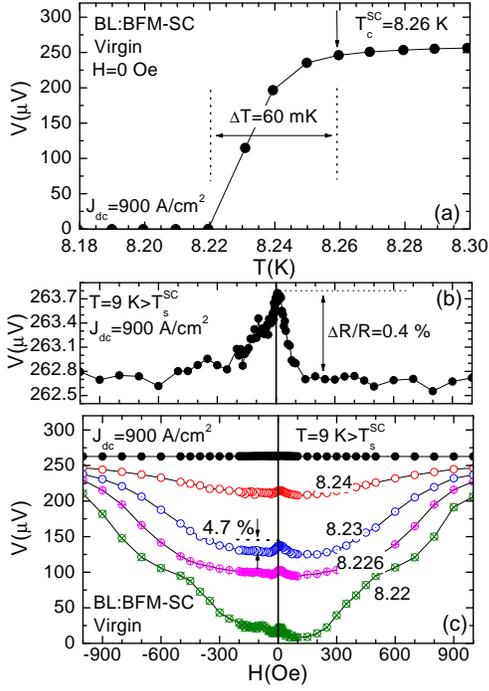}
\caption {(Colour online) (a) Zero-field resistive transition of a
BL:BFM-SC with T$_c^{SC}=8.26$ K. (b) Isothermal V(H) curve
obtained at T$=9$ K$>$T$_c^{SC}=8.26$ K. (c) Isothermal V(H)
curves obtained at temperatures allocated across the zero-field
isofield resistive V(T) curve presented in panel (a). All the data
were obtained when the BL was virgin.}
\label{b11}%
\end{figure}%

Until now all the data that have been presented referred to
BFM-SC-PFM TLs. We have clearly shown that in these structures
three parameters mainly control the magnetoresistance peaks that
are observed in the low-field regime, close to T$_c^{SC}$: (i) the
out-of-plane rotation of the outer FM layers' magnetizations,
subsequently motivating their magnetic interconnection, (ii) the
in-plane relative magnetization configuration of the outer FM
layers, and (iii) the mechanism of exchange bias that controls the
magnetic order of the BFM bottom layer. The investigation of more
simple BFM-SC BLs could provide important information on these
topics when contrasted to the results obtained in the BFM-SC-PFM
TLs since in BLs parameter (ii) is absent and parameter (i) is
weakened. In the present work we have made such a detailed
investigation. Representative transport data for a BFM-SC BL that
exhibits T$_c^{SC}=8.26$ K are shown in
Figs.\ref{b11}(a)-\ref{b11}(c). Panel (a) shows its zero-field
resistive V(T) curve along which the detailed isothermal V(H)
curves have been obtained. Panel (b) shows reference data obtained
in the normal state i.e. at T$=9$ K$>$T$_c^{SC}=8.26$ K. We see
that the obtained curve exhibits a single relatively broad peak
placed around zero field. In this field interval the virgin NiFe
layer reverses its magnetization (see Fig.\ref{b12} below). Panel
(c) presents the respective V(H) curves obtained at temperatures
allocated across the zero-field resistive V(T) curve presented in
panel (a). In qualitative agreement with the normal state V(H)
curve shown in panel (b) the V(H) ones of panel (c) exhibit one
single peak that is placed around zero field. Quantitatively, the
situation is different since the magnetoresistance peaks observed
in the superconducting (normal) state of the BLs amount to almost
$5\%$ ($0.4\%$) according to the definition
$(R_{max}-R_{min})/R_{nor}\times100\%$. We stress that the normal
state V(H) curve is presented once again in panel (c): in this
range the minor normal state peak ($\Delta
R/R_{nor}\times100\%=0.4\%$) is not even discerned.

Finally, we note that an oscillating-like behavior observed not
only in the virgin (see the curve obtained at T$=8.22$ K in
Fig.\ref{b11}(c)) but also in the BS (see Figs.\ref{b12}(c) and
\ref{b13}(c), below) V(H) curves near the bottom of the zero-field
resistive V(T) curve is not understood at the moment. Measurements
focused on this effect are on the way and will be discussed in a
forthcoming publication.

\begin{figure}[tbp] \centering%
\includegraphics[angle=0,width=6.5cm]{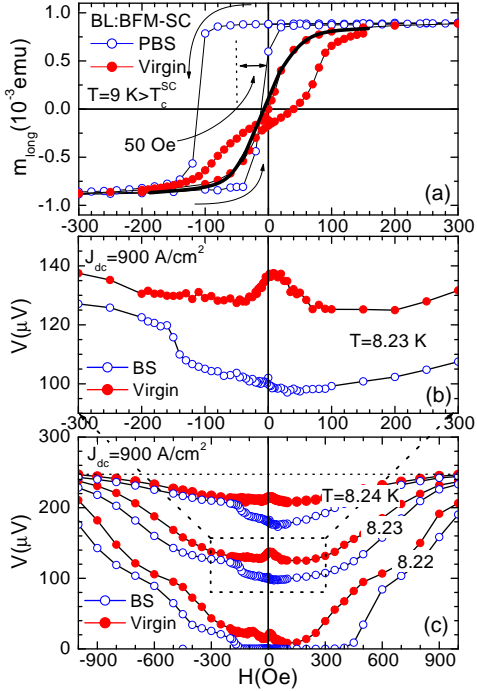}
\caption {(Colour online) (a) Detailed magnetization loops for
virgin (solid circles) and exchange biased (open circles) BL. (b)
Low-field part of a representative magnetoresistance V(H) curve
obtained at around the middle of the zero-field resistive
transition (see Fig.\ref{b11}(a)) when the BL was virgin (solid
circles) and when BS (open circles). (c) Detailed
magnetoresistance V(H) curves obtained just below the critical
temperature T$_c^{SC}=8.26$ K when the BL was virgin (solid
circles) and when BS (open circles) in an extended field regime.}
\label{b12}%
\end{figure}%

Until now the V(H) data presented in Fig.\ref{b11}(c) referred to
the case where the BL was virgin. A direct comparison between data
obtained when the BFM-SC BL is virgin (solid circles) and when BS
(open circles) is shown in Figs.\ref{b12}(a)-\ref{b12}(c). Panel
(a) shows detailed magnetization loops obtained at T$=9$
K$>$T$_c^{SC}$, while panels (b) and (c) show the respective
magnetoresistance data in the low-field and in an extended field
regime, respectively. Let us first discuss the magnetization data
shown in panel (a). A close inspection of the virgin loop reveals
that {\it when the positive part of the decreasing branch is
merged with the negative part of the increasing one a complete
typical loop of a soft FM is formed} (as presented by the thick
solid curve). Thus, we may conclude that when we decrease the
applied field from positive saturation it is the soft NiFe areas
that firstly reverse their magnetization and subsequently the hard
areas also reverse. In the same way when we increase the applied
field from negative saturation again the soft (hard) areas reverse
first (second). In contrast, when the NiFe layer is BS the two
different areas get strongly exchange coupled so that they reverse
simultaneously. In the decreasing branch, the point where the
abrupt switching of the strongly coupled soft/hard areas occurs in
the BS curve coincides with the field value where the gradual
reversal of the free/uncoupled {\it hard} phase is accomplished in
the virgin curve. In the increasing branch, the abrupt switching
of the BS curve coincides with the point of the virgin curve where
the magnetization reversal of the free/uncoupled {\it soft} phase
is completed. Thus, when the soft and hard areas are exchange
coupled it is the hard phase that determines the point where the
switching will take place for the preferential coupling direction
imposed initially by the external applied field, while the soft
phase determines the switching field in the opposite direction.

Now we may discuss the V(H) curves presented in panels (b) and
(c). The virgin V(H) curves present a minor peak as already
discussed above. Interestingly, this peak that originally is
centered around zero field may be shifted to negative field values
under the application of exchange bias. More importantly, we see
that while in the normal state the virgin and BS curves coincide,
as we enter in the superconducting state these curves diverge,
with the BS curve placed significantly below the virgin one in the
whole field range; a clear experimental proof that, apart from the
TLs, also in the BLs the exchange bias promotes the
superconducting properties of the whole hybrid. A clear difference
between the data obtained for the BLs
(Figs.\ref{b12}(a)-\ref{b12}(c)) and the TLs
(Figs.\ref{b5}(a)-\ref{b5}(c)) is the following: while in the BLs
the BS curves are placed below the virgin ones in the whole field
range, in the TLs there is a small field interval where the BS
V(H) curves overshoot the virgin ones. Most importantly, the
magnetoresistance peaks observed in the TLs are pronounced
[$(R_{max}-R_{min})/R_{nor}\times100\%=50\%$] when compared to the
ones occurring in the BLs
[$(R_{max}-R_{min})/R_{nor}\times100\%=5\%$]. We ascribe this
significant difference to the fact that the underlying mechanism
motivating this effect in the TLs is the magnetic coupling of the
outer layers through stray fields as the out-of-plane rotation of
their magnetizations takes place near coercivity. On the other
hand although in the BLs the out-of-plane rotation of the single
BFM layer's magnetization also takes place near coercivity the
mechanism of stray fields' interconnection is not possible.

\begin{figure}[tbp] \centering%
\includegraphics[angle=0,width=6.5cm]{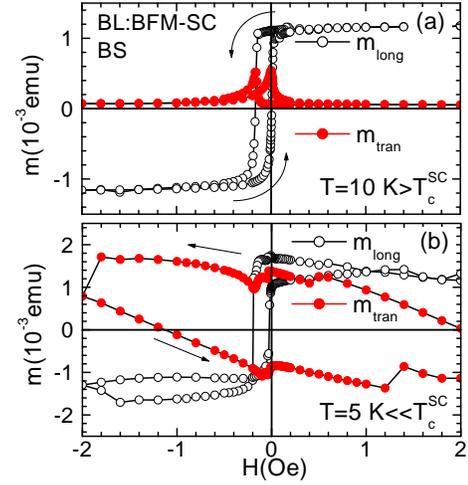}
\caption {(Colour online) Representative magnetization loops of
the BS BL's longitudinal (open circles) and transverse (solid
circles) components obtained (a) above T$_c^{SC}$ and (b) well
inside the superconducting state.}
\label{b18}%
\end{figure}%

However, the magnetization data presented in
Figs.\ref{b18}(a)-\ref{b18}(b) for another BL clearly demonstrate
that even in the BLs the out-of-plane rotation of the single BFM
layer's magnetization also occurs we believe that this fact
deserves a little more attention. This fact is not surprising
since the reversal of a FM layer's magnetization usually evolves
not only through the formation and movement of magnetic domains
but also through the in-plane or, as in our case the out-of-plane
rotation of the magnetization. A couple of interesting
observations that can be made from the presented data are the
following: First, in panel (a) we see that in the normal state the
transverse component's maximum (solid circles) occurs at exactly
the coercive field of the longitudinal one (open circles). Second,
in panel (b) we see that in the superconducting state neither the
longitudinal nor the transverse component exhibit the model loop
expected for a SC. In contrast, for the TL the transverse
component exhibited the fingerprint of a SC's model m(H) loop (see
the detailed loops presented in Figs.\ref{b16}(a)-\ref{b16}(b)).
We attribute this difference to the magnetic interconnection of
the outer FM layers that in the TLs forces the SC to behave
diamagnetically not in respect to the externally applied parallel
magnetic field but in respect to the transverse magnetization
field that emerges. Although in the BLs this exact mechanism is
missing the observed magnetoresistance peaks that are shifted from
zero field to negative coercivity (see Fig.\ref{b12}(c)) may also
be attributed to the out-of-plane rotation of the BL's
magnetization that occurs at exactly the same field value.

\begin{figure}[tbp] \centering%
\includegraphics[angle=0,width=6.5cm]{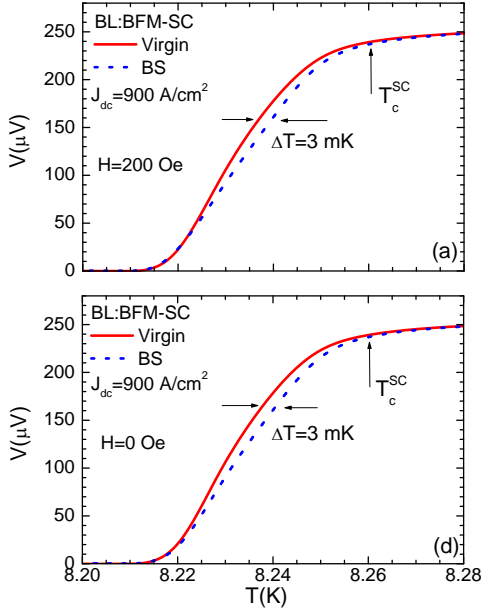}
\caption {(Colour online) Representative voltage curves as a
function of temperature at (a) $H_{ex}=200$ Oe and at (b)
$H_{ex}=0$ Oe, when the BL was virgin (solid lines) and when BS
(dotted lines).}
\label{b14}%
\end{figure}%

Except for the isothermal measurements as a function of magnetic
field the influence of exchange bias is also identified in the
isofield measurements as function of temperature. In
Figs.\ref{b14}(a) and \ref{b14}(b) we show representative data for
a BFM-SC BL for the cases when the BL was virgin (solid lines) and
when BS (dotted lines). Panel (a) shows the results obtained for
external field $H_{ex}=200$ Oe, while panel (b) for $H_{ex}=0$ Oe.
{\it We see that as we observed in the BFM-SC-PFM TLs also in the
BFM-SC BLs the BS state promotes superconductivity when compared
to the virgin one.} Also, these data clearly prove that the
observed effect is an exclusive property of the superconducting
state since in the normal state i.e. for T$>$T$_c^{SC}$ the
different resistance curves clearly coincide. Thus, the
enhancement of the superconducting properties of the BL can't be
attributed to a normal state property of the BFM layer that is
preserved in the superconducting state. It is a unique property
that is exclusively ascribed to the synergy of superconductivity
and exchange bias.

\subsection{Training effect in the superconducting state}

As it was discussed in subsection III.A where we presented a
complete magnetic characterization of the produced samples one
effect that is commonly observed in FM/AFM \cite{Gruyters00} and
soft/hard \cite{Sort04} exchange coupled systems is the so-called
training effect (see Fig.\ref{b3}(a)).\cite{Schuller99} According
to this effect in an exchange coupled system the shift of the
magnetization loop is decreased when successive loops are
performed without refreshing the coupling of the FM and AFM
constituents by appropriate field cooling from above the Neel or
blocking temperature.\cite{Schuller99} In Fig.\ref{b3}(a) we
presented two successive magnetization loops for a TL's
longitudinal component that were obtained above the
superconducting critical temperature.

\begin{figure}[tbp] \centering%
\includegraphics[angle=0,width=6.5cm]{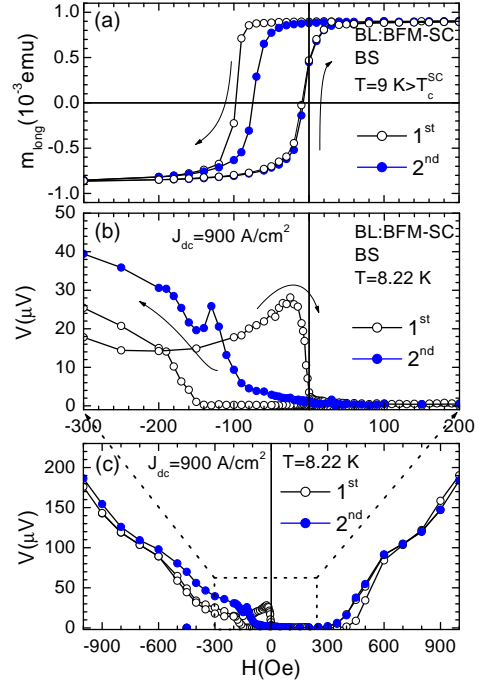}
\caption {(Colour online) (a) Two successive magnetization loops
obtained at temperature T$=9$ K$>$T$_c^{SC}$ in a BFM-SC BL.
Successive magnetoresistance V(H) curves obtained at T$=8.22$ K
near the bottom of its zero-field resistive V(T) curve (see Fig.
\ref{b11}(a)) in the low field (b) and in an extended field (c)
regime. All the data were obtained during a single run while the
BFM-SC hybrid was BS only once, prior to the first loop.}
\label{b13}%
\end{figure}%

In order to investigate if this effect also survives {\it inside}
the superconducting state we performed systematic transport
measurements. Figures \ref{b13}(a)-\ref{b13}(c) show
representative data obtained in a BFM-SC BL. Panel (a) shows two
successive magnetization loops obtained at temperature T$=9$
K$>$T$_c^{SC}$, while panels (b) and (c) show the respective
magnetoresistance V(H) curves obtained at T$=8.22$ K near the
bottom of its zero-field resistive V(T) curve (see Fig.
\ref{b11}(a)). Panel (b) focuses in the low-field regime, while
panel (c) shows an extended field range. We note that these data
were obtained when the BL was initially BS. In addition, these
data have been obtained successively i.e. without refreshing the
exchange coupling in the BFM NiFe layer. In panel (a) we see that
the width (W$=95$ Oe), and accordingly the shift, of the loop is
decreasing strongly during the second round (W$=75$ Oe). In
addition, in our case we observe that the first decreasing branch
is more square-like, while the subsequent one is more rounded.
This fact indicates that the underlying mechanisms motivating the
reversal of the BFM NiFe layer's magnetization during the first
and the subsequent loop is different. Moreover, we see that the
increasing branches clearly coincide in all successive rounds, a
sign of different magnetization reversal mechanisms between the
increasing and decreasing branches.

Panel (b) focuses on the V(H) data in the same field range as in
panel (a) in order to make a direct comparison between the
longitudinal magnetic component m$_{long}$(H) and the
magnetoresistance V(H) data. We see that during the first turn
(open circles) the whole hybrid is superconducting, since it
exhibits zero resistance, down to $-150$ Oe where the first
reversal of the NiFe layer's magnetization is completed. At this
field the V(H) curve gets non-zero since an abrupt increase is
observed for higher negative field values. The return V(H) branch
also presents an interesting behavior. While for negative magnetic
fields the BFM-SC hybrid superconducts only partially it is just
below zero field where it presents a clear peak and abruptly the
hybrid becomes completely superconducting since its resistance
vanishes at the field where the reversal of the NiFe layer's
magnetization is completed. Subsequently, the second decreasing
V(H) branch (solid circles) behaves qualitatively similar to the
first one (open circles) with the only quantitative difference
that it becomes partially superconducting at a lower field value
when compared to the first decreasing curve (the abrupt increase
occurs at $-90$ Oe and $-150$ Oe for the first and second rounds,
respectively). This is the fingerprint of the training effect that
according to our data survives in the superconducting state.

Panel (c) shows the V(H) curves presented in panel (b) in an
extended field range. We see that apart from a small interval of
positive magnetic fields the second decreasing V(H) curve (solid
circles) is placed above the first one (open circles). This means
that in our BFM-SC BLs the exchange bias is an essential
ingredient for maintaining the lower resistance that is possible
since the training effect weakens the exchange bias and
consequently the superconducting properties of the complete
hybrid.

\begin{figure}[tbp] \centering%
\includegraphics[angle=0,width=6.5cm]{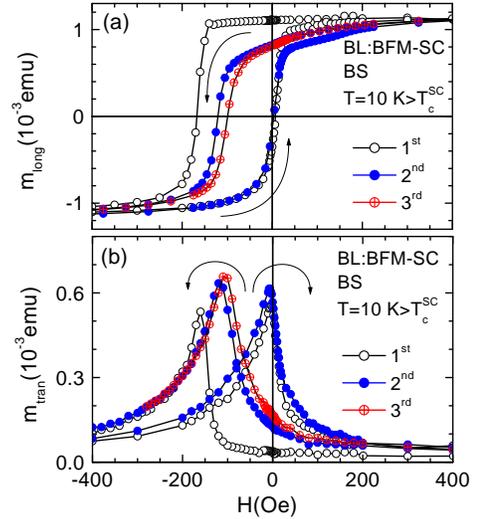}
\caption {(Colour online) Three successive magnetization loops for
the longitudinal (a) and transverse (b) components obtained at
temperature T$=10$ K$>$T$_c^{SC}$ in a second BFM-SC BL that
exhibits both higher coercive and exchange bias fields. All the
data were obtained during a single run while the BFM-SC hybrid was
BS only once, prior to the first loop.}
\label{b19}%
\end{figure}%

Since we ascribe the magnetoresistance peaks mainly to the
out-of-plane rotation of the magnetization the training effect
that is observed in the V(H) curves presented in
Figs.\ref{b13}(b)-\ref{b13}(c) should also be reflected in the
transverse magnetic component. Indeed, Figs.
\ref{b19}(a)-\ref{b19}(b) show detailed results for both the
longitudinal and transverse components for another BL that
exhibits slightly higher coercive and exchange bias fields. Panel
(a) shows three successive loops for the longitudinal component,
while panel (b) presents the respective data for the transverse
one. Once again we stress that in these measurements the BL was
exchange biased by cooling it above its blocking temperature
T$_B=100$ K under the presence of a positive magnetic field only
once, prior to the very first measurement. A number of notes may
be drawn from these data. First, we clearly see that the first
loop (open circles) of the longitudinal [transverse] component is
significantly asymmetric with the decreasing branch exhibiting a
sharp downfall [rise] at $150$ Oe. Second, the first loop of the
longitudinal component is more square-like in comparison to the
two subsequent ones. Third, the maximum value of the transverse
component increases progressively with the number of performed
loops. Fourth, both the coercive and exchange bias fields decrease
as the number of loops progressively increases. More importantly,
the decrease of these two parameters (coercive and exchange bias
fields) is abrupt during the first two loops, while the third
subsequent loop has only a minor effect. For instance, the full
width is W$=170$ Oe for the first round, while it gets W$=120$ Oe
and W$=100$ Oe in the second and third rounds, respectively. This
behavior is very similar to the one obtained by Dieny and
colleagues \cite{Sort04} in NiFe/[Pt-Co] soft/hard exchange
coupled BLs where it was observed that the reduction of the
longitudinal component loop's width is particularly steep during
the first three rounds.\cite{Sort04} Here, we reveal that the same
behavior also holds for the transverse magnetic component. All
these facts are direct consequences of the training effect: {\it
under subsequent loops the initially induced exchange bias is
weakened so that the in-plane magnetic order can't be maintained
and the out-of-plane magnetic component progressively increases}.
These facts are also imprinted on the magnetoresistance curves
presented in Figs.\ref{b13}(b)-\ref{b13}(c). We clearly see that
the morphology of the transverse component's m$_{tran}$(H) curves
presented in Fig.\ref{b19}(b) is identical to the one of the
magnetoresistance V(H) curves shown in Fig.\ref{b13}(b). The
training effect that is presented here for the BLs was also
observed for the TLs.

Once again, we underline that although in the BLs it is also the
transverse magnetic component that attains significantly high
values near coercivity the observed magnetoresistance peaks are
only minor when compared to the ones observed in the TLs. This is
why we believe that the underlying mechanism motivating the TLs'
extreme magnetoresistance peaks is the magnetic coupling of the
outer FM layers through the emergence of magnetic domains and the
accompanying stray fields that occur near
coercivity.\cite{Gider98,StamopoulosRCPRBSub,Parkin00} In the BLs
the single BFM layer doesn't have this opportunity; owing to the
absence of a second layer a complete stray-fields magnetostatic
coupling can't be accomplished.

\section{Summary and Conclusions}

Summarizing, in this work we studied in detail the magnetic and
transport properties of exchange biased TLs and BLs that
constitute of low spin polarized Ni$_{80}$Fe$_{20}$ and low-T$_c$
Nb when the external magnetic field was applied parallel to the
specimens. We observed the following experimental facts:

(i) The mechanism of exchange bias may relatively promote
superconductivity. This is revealed by a clear positive shift of
the isothermal V(H) and isofield V(T) curves when the hybrids are
exchange biased. In addition, the presented I-V characteristics
prove that the current-currying capability of the hybrids is
significantly improved when they are exchange biased.

(ii) The fingerprints of exchange bias, that is the unidirectional
anisotropy that results in a shift in the magnetization loops and
the training effect that relates to weakening of the mechanism
upon performing sequential loops are clearly reflected in the
magnetoresistance curves that are obtained in the {\it
superconducting} state.

(iii) The observed magnetoresistance peaks are major for the
BFM-SC-PFM TLs [$(R_{max}-R_{min})/R_{nor}\times100\%=50\%$], but
are only minor for the BFM-SC BLs
[$(R_{max}-R_{min})/R_{nor}\times100\%=5\%$]. We ascribe this
important difference to the fact that the underlying mechanism
motivating this effect in the TLs is the magnetic coupling of the
outer FM layers through stray fields as the out-of-plane rotation
of their magnetizations takes place near coercivity. These stray
fields may exceed, primarily the SC's lower critical field, or,
secondarily its upper one. In the first case the dissipation is
motivated owing to flow of vortices under the driving Lorentz
force that is exerted by the applied transport current, while in
the second case the normal areas that are {\it locally} formed
should contribute extra dissipation. Although in the BLs the
out-of-plane rotation of the single BFM layer's magnetization also
takes place the opportunity of intense stray-fields coupling is
not available due to the lack of a second FM layer so that the
magnetoresistance effect is only minor.

(iv) Since the exchange bias controls the in-plane magnetic order
it also controls the out-of-plane rotation and the related
stray-fields magnetic coupling of the outer FM layers in the TLs.
Thus, by means of exchange bias the current-carrying capability of
a TL "spin valve" (see Fig.\ref{b22}) could be tailored at will.
However, we should stress that in such "spin valves" the
underlying "valving mechanism" doesn't rely on the sophisticated
spin-dependent filtering process that was theoretically proposed
in Refs.\onlinecite{Buzdin99,Tagirov99} but on more conventional
pair-breaking mechanisms. Thus, in our case such a TL should be
rather called "supercurrent switch". Practical convenience is
readily realizable since such devices could switch between the
normal and superconducting states under the application of
exchange bias on the outer FM electrodes.

(v) Going a step farther we propose that a generic prerequisite
for the occurrence of extended and intense magnetoresistance peaks
in a FM-SC-FM TL is that the outer FM layers should have almost
the same coercive fields. When this condition is fulfilled these
layers are susceptible to magnetic coupling since magnetic domains
will occur simultaneously in their whole surface. Consequently,
the accompanying stray fields that naturally emerge above domain
walls will accomplish the magnetic coupling. Since in our TLs the
one FM layer incorporates the mechanism of exchange bias its
coercive field is tunable, while the respective coercivity of the
other layer is fixed. Thus, by tuning the coercive field of the
one FM layer we may satisfy the prerequisite of equal coercive
fields for the occurrence of broad and intense magnetoresistance
peaks.

The observations discussed right above may have a direct impact on
the interpretations made in recent works that have dealt with
relevant topics.

(i) The data presented in Figs. \ref{b13}(a)-\ref{b13}(c) and
\ref{b19}(a)-\ref{b19}(b) are important when recent experiments
that were performed in exchange biased FM-SC-FM TLs are considered
(see
Refs.\onlinecite{Gu02,Potenza05,Moraru06,Moraru06B,StamopoulosPRLSub}).
Such experiments should follow a strict experimental protocol. For
instance, consider the case when successive isothermal
magnetoresistance V(H) curves are measured as function of magnetic
field at different temperatures in a TL that was exchange biased
only once, prior to the first measurement. Although the original
indention was to get information on the magnetic field dependence
of the TL's transport behavior when this was exchange biased in
every case, due to the training effect the only reliable
information refers to the isothermal magnetoresistance V(H) curve
that is obtained during the first round of the experiment. All the
subsequent ones provide an {\it underestimation} of the TL's
current-currying capability since the training effect weakens
significantly the exchange bias that was imposed originally prior
to the very first measurement. Thus, experimental works where the
exchange bias has been employed should be reconsidered under the
light of these characteristics. Strict experimental protocols
should always be used when the transport behavior of such exchange
biased TLs and BLs is studied otherwise their properties may be
underestimated.

(ii) All experimental works that treat FM-SC-FM TLs, whether these
are plain or exchange biased (see
Refs.\onlinecite{Gu02,Potenza05,Moraru06,Moraru06B,Pena05,Rusanov06,Visani07,Singh07,Steiner06,StamopoulosPRLSub})
and are considered as exact realizations of the theoretical
propositions that refer to the "spin valve" concept (see
Refs.\onlinecite{Buzdin99,Tagirov99}) should be reexamined under
the light of the results presented in this work for exchange
biased TLs and of those presented in
Ref.\onlinecite{StamopoulosRCPRBSub} for plain ones. Probably, in
all cases an out-of-plane rotation and subsequent stray-fields
induced magnetic coupling of the outer FM layers could be involved
so that their in-plane relative magnetic configuration is not the
dominant mechanism that motivates the observed effects. If this is
true the theoretical propositions made in
Refs.\onlinecite{Buzdin99,Tagirov99} still wait for experimental
evidence.

(iii) Except for results obtained in TLs also the ones that refer
to more simple FM-SC BLs, whether they are plain or exchange
biased (see
Refs.\onlinecite{Rusanov05,Steiner06,StamopoulosPRLSub}) should be
treated more thoroughly. As it was clearly shown in this work for
exchange biased BLs and in Ref.\onlinecite{StamopoulosRCPRBSub}
for plain ones even in such simple structures the out-of-plane
rotation of the FM layer's magnetization could motivate a minor
magnetoresistance effect and falsify the obtained experimental
results. For instance, in Ref.\onlinecite{Rusanov05} plain NiFe-Nb
BLs were studied by transport measurements. The increase that was
observed \cite{Rusanov05} in the magnetoresistance curves prior to
the clear dips that were ascribed to the formation of "domain wall
superconductivity" \cite{Rusanov05,StamopoulosPRB06,Buzdinnew}
could be related to a partial out-of-plane rotation of the NiFe
layer's magnetization.

Finally, our data show clearly that the mechanisms of exchange
bias and superconductivity, that generally are considered so
fundamentally different under specific circumstances may even
become cooperative. In our work we used only a phenomenological
basis of the involved phenomena in order to give consistent
explanations for our experimental results. We hope that our
experimental work not only will shed light on the contradictions
that have been reported in the recent literature but will
eventually trigger new experimental and theoretical work that will
deal with the microscopic implications in these topics. Apart from
their importance for basic physics our results could also assist
the design of practical devices where the current-carrying
capability of a FM-SC hybrid could be efficiently controlled by
the exchange bias mechanism.

\pagebreak

\end{document}